\documentclass{aa} 

\usepackage{graphicx}
\begin{document}

   \title{The temporal and spatial evolution of the starburst in ESO\,338-IG04 
as probed by its star clusters
\thanks{Based on observations with the NASA/ESA {\it Hubble Space 
Telescope}, obtained at the Space Telescope Science Institute, which is 
operated by the association of Universities for Research in 
Astronomy, Inc., under NASA contract NAS5-26555.}}

\author{G\"oran \"Ostlin \inst{1}
	\and
	Erik Zackrisson \inst{2}  
	\and
	Nils Bergvall   \inst{2}      
	\and
	Jari R\"onnback \inst{2} }

\offprints{G\"oran \"Ostlin}

\institute{
	Stockholm Observatory,  SE-106 91 Stockholm, Sweden \\
	\email{ostlin@astro.su.se}
	\and
	Uppsala University, Dept. of Astronomy and Space Physics, Box 515, SE-75120 
	Uppsala, Sweden }

   \date{Received 0000; accepted 0000}

   \abstract{
\object{ESO 338-IG04}, also known as \object{Tololo 1924-416}, is a well known luminous 
 blue compact galaxy in the local universe. 
 Images obtained with the Hubble Space Telescope (HST) have shown that the central 
starburst region 
 is composed of numerous bright point sources - young star clusters, surrounded by 
 a population of old and intermediate age globular clusters. 
In this paper we use ultra-violet (UV) and optical HST photometry in five bands, and an extensive 
set of spectral evolutionary synthesis scenarios to investigate the age and masses of 
124 star clusters. The very small reddening makes  ESO\,338-IG04 an excellent laboratory for 
studying the formation of such objects. We find that a careful treatment of the 
nebular emission component is crucial when modelling the broad-band colours of young
starburst regions.
We have used the star clusters to trace the temporal and spatial evolution
of the  starburst, and to put constraints on the star formation activity
over a cosmological time-scale. The present starburst has been active for about
40 Myr and shows evidence for propagating star formation and structures triggered
by  galactic winds. A standard Salpeter initial mass function (IMF) 
extending up to 120 
$M_\odot$ provides the best fit to the data, although a flatter IMF cannot be
excluded. The compact star clusters provide 30-40\% of the UV luminosity and star
formation activity. We find no evidence for dust obscuration even among the youngest
($< 1$ Myr) clusters, and we propose that this may be related to a short time-scale
for destruction of dusty molecular clouds. Over a longer time-scale, we find evidence for
previous cluster formation epochs - notably one a couple of Gyr ago.
The fraction of the galaxy's stellar mass contained in compact star clusters 
is found to be several percent, which is an unusually high value. The intermediate
age clusters show a flattened space distribution which agrees with the isophotal
shape of the galaxy, whereas the oldest clusters seem to have a spherical distribution
indicating that they formed prior to the rest of the galaxy. 

      }

\authorrunning{\"Ostlin et al.}

\titlerunning{The Starburst in ESO\,338-04}

\maketitle

\section{Introduction}

According to the hierarchical paradigm for structure formation,
 small structures form first and massive galaxies are successively
built up by mergers of smaller units.
Blue Compact Galaxies (BCGs) are galaxies with lower mass and metallicity than 
typical $L^\star$ galaxies. Their spectra often show a blue continuum and 
strong emission 
lines, indicative of low chemical abundances and active star formation. 
These properties once led to the idea that BCGs were young systems presently 
forming their first generation of stars (Searle \& Sargent \cite{ss}).
Today it is evident that the majority of BCGs are not young (see Kunth \& 
\"Ostlin \cite{kunthostlin} for a review), but rather experiencing a 
phase of increased star formation activity, ranging from moderate enhancements 
(Schulte-Ladbeck et al. \cite{schulte}, Hopkins et al. \cite{hopkins}) to  
true starbursts, with short gas consumption time-scales (\"Ostlin 
et al. \cite{ostlin2001}). Understanding the reasons for this  lies at the 
heart of the quest to understand the role of BCGs in the cosmic galaxy evolution 
framework.

Observations of  e.g. the Hubble Deep Field (Williams et al. 
\cite{hdf}), show that luminous compact blue galaxies 
are common at higher redshifts (Guzman et al. \cite{guzman}). 
Moreover,   Ly$\alpha$ emitters at 
redshift $z=2.4$ have lower continuum luminosity and smaller sizes 
than expected for young $L^\star$ galaxies (Pascarelle et al. \cite{pascarelle}).
These may be dwarf
galaxies, about to build up larger
structures by mergers. Several investigations (e.g. Lilly et al. \cite{lilly},
Madau et al. \cite{madau}) show that the cosmic star formation rate (SFR) 
was much higher at $z\ge 1$, and
the same applies to the merger frequency although mergers appear to have
played an important role also at $z<1$ (e.g. Le F\`evre et al. \cite{lefevre}). 
Based on  kinematics, morphology and photometry, \"Ostlin et al. 
(\cite{ostlin2001}) suggest that mergers may be the main mechanism 
responsible for starbursts in luminous BCGs (LBCGs) in the local universe.
This is also supported by the enigmatic under-abundance of the ISM
metallicity compared to the stellar population found by Bergvall \& 
\"Ostlin (\cite{bergvallostlin}). Hence, LBCGs offer an opportunity to study  
hierarchical buildup of galaxies in detail.

Globular clusters (GCs) include the oldest known stellar
 systems in the local universe. Old Galactic halo GCs are around 15 Gyr old and 
typically have metallicities [Fe/H]$=-1.5$ (Ashman \& Zepf \cite{ashmanzepf}). 
High resolution imaging, notably with the Hubble Space Telescope (HST)
 of merging galaxies in the local universe have shown that luminous compact star 
clusters (often referred to as super star clusters) with properties consistent 
with  young
GCs are formed in such events (see 
Whitmore \cite{whitmore_rev} for a recent review). 
This can explain the existence 
of intermediate age (a few Gyr) metal-rich GC populations found in many local
elliptical galaxies (see Ashman \& Zepf \cite{ashmanzepf} for a review). 
Hence, GC formation was not confined to the 
high $z$ universe, but is still going on but at a slower pace. 
Also LBCGs are sites of young compact star clusters (\"Ostlin \cite{ostlin2000}).
However, the ISM abundances ([O/H]$\approx -1$) are  lower than 
those  in giant mergers, and similar (if [Fe/O]$\approx -0.5$, 
as is common in metal-poor environments, see e.g. Kunth and \"Ostlin 
\cite{kunthostlin}) to the metallicity of old GCs. 
The first generation of GCs in the universe may 
have formed   in the 
processes of merging sub-galactic clumps, similar to what is going on in some local LBCGs.

ESO\,338-IG04 is a well known luminous $(M_V=-19.3)$ BCG, first studied in 
detail by Bergvall (\cite{bergvall}). 
Images obtained with the
Wide Field Planetary Camera 2 (WFPC2) on board the HST
found numerous compact star clusters associated with ESO\,338-IG04 (\"Ostlin
et al. \cite{ostlin1998}, hereafter \"OBR98). In addition UV observations with  
the Faint Object 
Camera (FOC) on board HST, had already shown that the centre of ESO\,338-IG04 
hosted a population of UV-bright  star clusters, just like many other
starbursts (Meurer et al. \cite{meurer}). Further analysis showed that the
star clusters in ESO\,338-IG04 spanned ages from a few Myr to more than 10 Gyr
and that they had just the properties expected for young and old GCs (\"OBR98). 
The galaxy is a true starburst in the sense that the time-scales for
gas consumption and build-up of the observed stellar mass with the current
SFR both are $\sim 1$ Gyr, i.e. much shorter than the age of the universe
(\"Ostlin et al. \cite{ostlin2001}).

In the present paper, we will make a refined analysis of 124 star clusters in 
ESO\,338-IG04 using additional UV data and a new set of models. We use the star clusters 
and their age- and  space-distributions  to track the evolution of the star formation
in the galaxy over a cosmological time scale.

\begin{figure*}
\caption{{\bf See separate jpg-file: 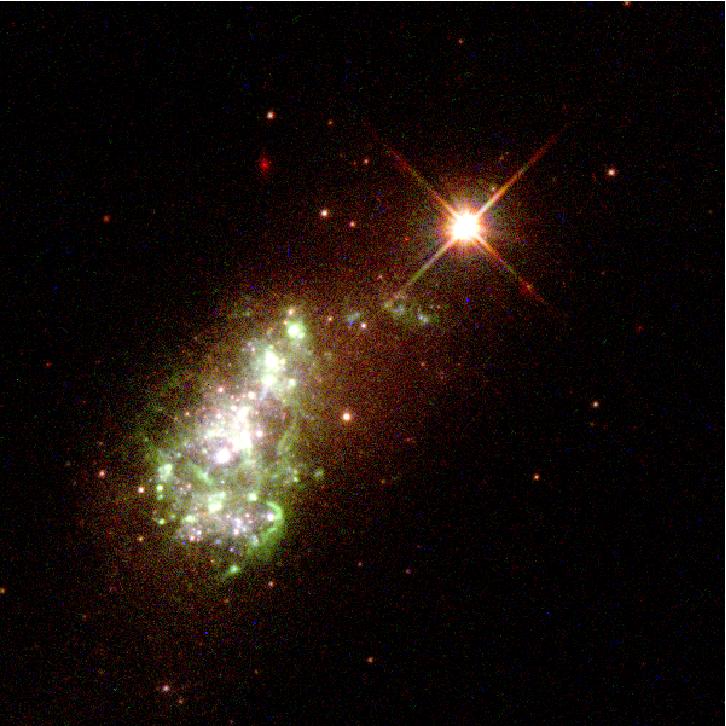}
True colour image composite of the central $30\arcsec \times 30\arcsec$ 
(corresponding to $5.45 \times 5.45$ kpc at the adopted distance 37.5 Mpc) region
of the Planetary Camera. F814W ($i$) is coded in red, F555W ($v$) in green and F439W($b$) in blue. North
is up-left (65 degrees measured from vertical towards left) and east is down-left. The 
bright source with diffraction spikes $\sim 5\arcsec$ from the centre towards the upper 
right is a Galactic foreground star.}
\label{image}
\end{figure*}

\section{Observational data}
 
\object{ESO 338-IG04} was observed with HST/WFPC2  with 
the galaxy centred on the Planetary
Camera (PC) aperture (\"OBR98).
The photometry in the F336W, F439W, F555W and F814W bands (hereafter  
referred to as: $u, b, v, i$) is described in \"OBR98. In addition we use 
photometry in the F218W filter  which is very valuable for constraining 
the physical parameters and the extinction of young objects. 
The F218W photometry (hereafter  $x$) has been performed in the same 
way as for the other passbands, and has been corrected for the
same amount of Galactic extinction: $E(B-V)_G =0.09$ (Burstein and Heiles 1982).
We have also corrected the F218W data for a ``UV-contamination'' of $-0.1276$ magnitudes
(Biretta et al. \cite{biretta}). All photometry in this paper is in the 
VEGAMAG system (Biretta et al. \cite{biretta})

\begin{figure}
\caption{{\bf See separate jpg-file: 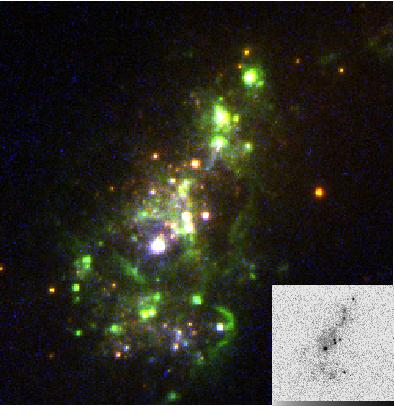}
True colour RGB composite of the central $11.6\arcsec \times 11.6\arcsec$
($2 \times 2$ kpc) region with cuts chosen to show details of the central bright regions:
F814W ($i$, red), F555W ($v$, green)  and F218W($x$, blue). The greyscale inset shows
the F218W image in of the same field.}
\label{image2}
\end{figure}

Figure \ref{image} shows an RGB true colour composite of the $i$, $v$ and 
 $b$ images. The  $v$  filter (F555W) has significant transmission at H$\beta$, 
[\ion{O}{iii}]$_{\lambda 4959,5007}$
and even H$\alpha$. Consequently, several nebulous emission-line regions 
are seen in green in Fig. \ref{image}. 
It is obvious  that this galaxy presents a large number 
of compact star cluster candidates. In  \"OBR98 we showed that the vast majority
of these are indeed star clusters, physically associated with ESO338-IG04 (i.e. 
not foreground stars, background galaxies or individual supergiants in ESO338-IG04).
Their sizes are compatible with globular clusters, although
a few of the younger sources are slightly more extended and may represent
 unbound systems. In fig. \ref{image2} we show an RGB composite of the central
region using the $i, v$ and $x$ filters. 
The variety of colours evident from Figs. \ref{image} and \ref{image2}: 
a central population with bright blue clusters
surrounded by a population of fainter red ones,   
directly suggests that the clusters have different age or reddening.

\begin{figure}
\resizebox{\hsize}{!}{\includegraphics{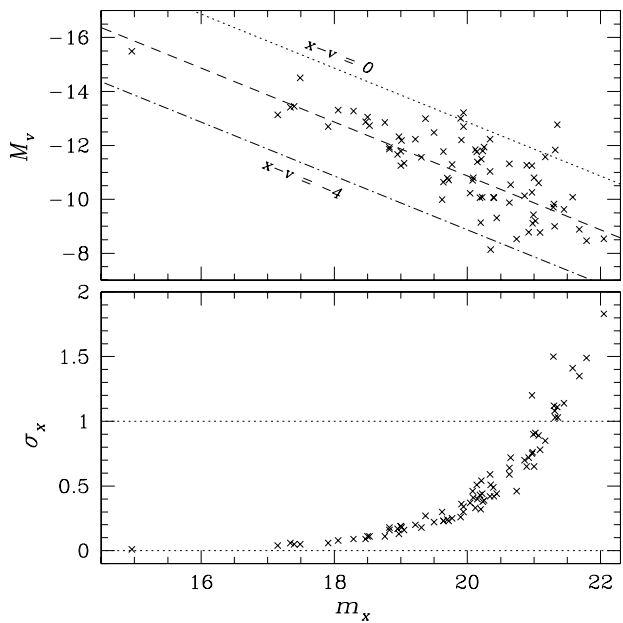}}
\caption{Photometry for clusters detected in the F218W ($x$) band. The bottom panel 
shows the  magnitude--uncertainty diagram. Top panel shows
the apparent $x$ magnitude vs the absolute $v$ band magnitude.
 Lines of constant colour are 
shown as well: $x-v = 0$ (dotted), $-2$ (dashed), and $-4$
(dash-dot).}
\label{magmag}
\end{figure}

Figure \ref{magmag} shows the F218W magnitude ($m_x$) vs the photometric 
uncertainties ($\sigma_x$), and vs $M_{v}$,
the absolute $v$ magnitudes for the same distance as used in \"OBR98
(37.5 Mpc, or $m - M = 32.87$)\footnote{The distance is based on a Hubble parameter of H$_0 =  
75$ km/s/Mpc, which is assumed throughout this paper.}. 
From the catalogue in  \"OBR98, a total of 78 objects 
with  $m_x \le 22$ were detected. 
However,  cosmic ray contamination is worse (due to longer 
exposure time, 1800s split into two sub-exposures) than in the other filters,
and some spurious detections had to be removed by hand. In total,
49 objects have $\sigma_x \le 0.5 $ and 21 objects have $\sigma_x 
\le 0.2$.
This is comparable to Meurer et al. (\cite{meurer}) who detected 14 compact 
sources with the pre-refurbishment HST/FOC and the F220W filter having uncertainties 
$\sigma_{F220W} \le 0.17$. The photometric agreement is fair when 
comparing sources for which Meurer et al. (\cite{meurer}) determined 
the flux from  radial profiles.   
From  Fig. \ref{magmag} it is evident that the sources detected in F218W
have very blue $x-v$ colours, with
a median around $-2$, a combined effect of low age, metallicity and extinction
of the UV detected clusters.

\begin{figure}
\resizebox{\hsize}{!}{\includegraphics{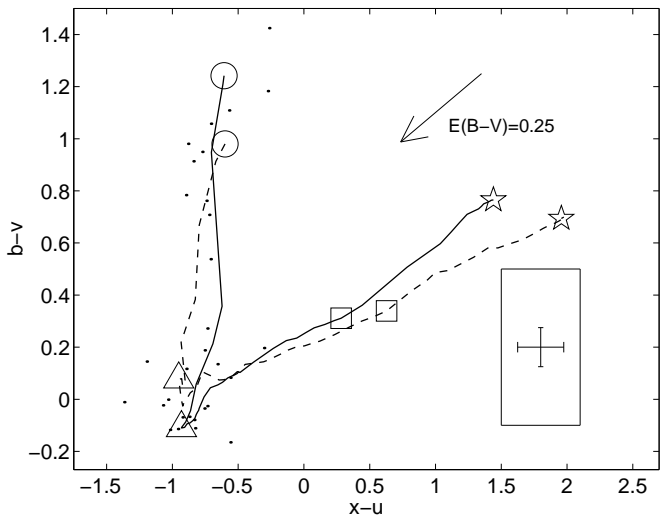}}
\caption{Colour-colour diagram including the F218W data (denoted 'x').
 The median colour uncertainty 
for objects included in the plot is shown as a cross in the lower right, 
whereas the largest uncertainty associated with any object is indicated 
by the surrounding rectangle. 
The arrow indicates how much and in what direction the objects would be 
repositioned after correction for a hypothetical  extinction of 
$E(B-V)=0.25$. The
predictions for two of the used spectral evolutionary synthesis 
models (Salpeter IMF, $M = 0.08$ to $120 M_\odot$, instantaneous burst, 
$Z_{\rm gas} = 0.002$, $Z_{\rm stars} = 0.001$) are shown as well: 
Zackrisson et al (2001, solid line), and P\'EGASE.2 (dashed line). Markers 
indicate ages of 0.5 Myr (circles), 10 Myr (triangles), 
1 Gyr (squares) to 15 Gyr (pentagrams).}
\label{xubv}
\end{figure}

Figure \ref{xubv} shows the $x-u$ vs $b-v$ two-colour diagram 
for objects with accurate photometry,
together with two typical models used (see below for details on these).
During the first 10 Myr the $x-u$ colour is nearly constant making it an 
efficient probe of the reddening for young objects. On the other hand, 
$b-v$ evolves strongly, mainly due to the rapid decrease in the contribution
from emission lines to the $v$ (F555W) filter. That the objects scatter close
to the tracks demonstrate that the extinction is very small. 
It is also clear that it is only for
objects younger than a few times 10 Myr that we have reliable F218W data.

In addition to the clusters in Fig. \ref{image}, which are found in the 
unvignetted field of view (FOV) of the PC, a blue tail 
of the galaxy (see Bergvall \& \"Ostlin \cite{bergvallostlin}) extends into 
the WF4 aperture (downwards in the orientation of  Fig. \ref{image}), 
where another dozen compact objects are found. This is near the 
border between the chips, and we do not regard the 
photometry of these objects reliable. Some objects appear extended, others
resemble compact star clusters, which is confirmed by our new deep STIS data
(\"Ostlin et al. in prep.). The STIS images (FOV $9.3 \times 9.3$ kpc) also show 
a small number of 
bright clusters  lying outside the upper and right boundaries of the PC FOV. 
Bergvall and \"Ostlin trace the luminosity profile 
of ESO\,338-04 to radii greater than 10 kpc, and in \"OBR98 it was shown that the radial
cluster distribution follows that of the visual light.  From this we estimate that
the limited area of the PC results in a  cluster incompleteness, at 
the $\sim 11$\% level, in addition to the photometric incompleteness. 

In Table \ref{starburst} we present the integrated starburst luminosity in
different passbands and the fraction of light attributable to compact star clusters.

\section{Spectral evolutionary synthesis models}
Due to the uncertainties still prevailing the field of spectral 
evolutionary synthesis, all results
 should ideally be cross-checked with independent models. To 
ensure that the age and mass distributions derived for 
our star cluster sample do not critically depend on the predictions 
from any single code, the analysis has been carried out with two models 
in parallel:  P\'EGASE.2 (Fioc \& Rocca-Volmerange \cite{pegase}) 
and the model by Zackrisson et al. (\cite{Zackrisson 
et al.}, hereafter Z2001). These particular models were chosen since they contain
a nebular component and because of the ease with which non standard filters may be included. 

P\'EGASE.2 is based on the method of isochrone synthesis, employs 
stellar atmospheres by  Lejeune (\cite{Lejeune et al.}) and Clegg 
\& Middlemass (\cite{Clegg & Middlemass}), together with 
stellar evolutionary tracks mainly from the Padova group. Using a 
simple prescription for the nebular component, it predicts both 
continuum and line emission of the ionised gas. 

Z2001 is based on the method of isomass synthesis, employs the same 
set of stellar atmospheres as P\'EGASE.2 but stellar evolutionary 
tracks mainly from the Geneva group. Pre-main sequence evolution and 
a stochastic treatment of horizontal branch morphologies 
at low metallicities are also included. Gas continuum and emission lines are predicted using 
the photoionisation code Cloudy version 90.05 (Ferland \cite{Ferland}).
For each time step, the spectral energy distribution of the 
stellar population is fed into Cloudy, making the computation 
time consuming, but also giving a more realistic nebular emission 
component.

The use of pre-main sequence evolution and and a more sophisticated nebular 
component in Z2001 makes this model more suited for the current analysis, 
where many clusters are believed to be young. As will be demonstrated 
(see Sect. 4.1, 5.1) the  predictions by Z2001 are also in better agreement 
with the observations. A large grid of evolutionary 
sequences was therefore generated using Z2001, and a smaller set, intended 
to use as a consistency check, with P\'EGASE.2. The details of these 
sequences are given in tables 1 and 2. In total, 133 parameter 
configurations were processed with Z2001, 10 with P\'EGASE.2.
The two models used represent an improvement over the one used in \"OBR98 in terms of 
the nebular component, time resolution at young ages, updated stellar 
evolutionary tracks and stellar 
atmospheres at a wider range of metallicities.
In addition, the Z2001  grid represents an increase in the size of
the parameter space by 
more than a factor of 30 compared to  \"OBR98.

Magnitudes in the HST Vegamag system were generated from model spectra
using the HST/WFPC2 filter+CCD response curves, and a theoretical Vega 
model. Since the full transmission profiles were
used, red leaks are included in the calculation and do not result in
any degenercies. The stellar initial mass function 
(IMF) was parametrised as a power-law: 
$dN/dM \propto M^{-\alpha}$, with spectral index $\alpha$. For a Salpeter
(\cite{salpeter}) IMF, $\alpha=2.35$. 
The Scalo (1998, hereafter Scalo98) IMF used in some of the P\'EGASE.2
evolutionary sequences was parametrised by a piecewise continuous 3 segment power-law.

\begin{table}[h]
\caption[]{The grid of Z2001 (Zackrisson et al. \cite{Zackrisson et al.}) evolutionary 
sequences. The grid consists of all 
possible combinations of the parameter values listed below. To furthermore 
ensure the robustness of the final results, a smaller number of evolutionary sequences was 
also calculated with variations like:  
constant star formation during 5 Myr and 100 Myr, $M_{\mathrm{up}}=40, 
\ 60, \ 80 M_\odot$, $10^6 M_\odot$ available for star formation, $n(\mathrm{H})=100 \
 \mathrm{cm^{-3}}$, filling factor 1.0, $Z_{\mathrm{stars}}=0.004, \ 0.008, 
\ 0.020, \ 0.040$ and $Z_{\mathrm{gas}}=0.001, \ 0.004, \ 0.008, \ 0.020, \ 0.040$. 
Neither the variations in hydrogen density, filling factor nor stellar mass were seen 
to have any serious impact on the final ages and masses derived.}
\begin{flushleft}
\begin{tabular}{ll} 
\hline
IMF, $\alpha$ & 1.35, 2.35, 2.85 \cr
$M_{\mathrm{low}}$ & 0.08, 2 $M_\odot$\cr
$M_{\mathrm{up}}$  & 20, 120 $M_\odot$\cr
SFH & c1e7, e3e6, Inst \cr
$Z_{\mathrm{stars}} $ & 0.001 \cr
$Z_{\mathrm{gas}} $   & 0.002 \cr
$M_{\mathrm{stars}} $ & $10^5 M_\odot$ \cr
$n(\mathrm{H}) $& 10 cm$^{-3}$\cr
Filling factor & 0.1 \cr
Covering factor & 0.5, 1.0\cr
Nebular emission & Yes, No\cr
\hline
\end{tabular}\\
IMF: $dN/dM \propto M^{-\alpha}$\\
SFH=Star formation history: c1e7 = Constant SFR during 10 Myr,
e3e6 = Exponentially declining SFR  with e-folding decay rate of 3 Myr,
Inst = Instantaneous burst.
\label{z2001}
\end{flushleft}
\end{table}

\begin{table}[h]
\caption[]{List of used P\'EGASE.2 evolutionary sequences,  all assuming 
$Z_{\mathrm{stars}}=Z_{\mathrm{gas}}=0.002$. 
All evolutinary sequences were
computed both with and without inclusion of nebular emission.} 
\begin{flushleft}
\begin{tabular}{llll} 
\hline
IMF & $M_{\mathrm{low}} (M_\odot)$ & $M_{\mathrm{up}} (M_\odot)$ & SFH\cr
\hline 
Salpeter & 0.1 & 120 & Inst\cr
Salpeter & 0.1 & 20 &  Inst\cr
Salpeter & 0.1 & 120 &  e3e6 \cr
Salpeter & 1 & 120 &  Inst\cr
Scalo \cite{Scalo} & 0.1 & 120 &  Inst \cr
\hline
\end{tabular}
\end{flushleft}
\end{table}

\begin{table}[h]
\caption[]{Integrated starburst  luminosities. All values have been corrected 
for foreground reddening by our Galaxy. The first
column gives the name of the filter, and the second column the effective
wavelength of the filter+CCD combination (Biretta et al. 2002). 
The third  column gives the integrated starburst magnitude 
(in the VEGAMAG system) within a radius of 5.9\arcsec (1.075 kpc) centred on 
the brightest cluster. The fourth column gives the same quantity
in $f_\nu$ units. The fifth column gives the magnitude of the
starburst region after all detected compact clusters have been subtracted. The
last column gives fractional contribution from  compact star clusters 
to the integrated starburst luminosity.} 
\begin{flushleft}
\begin{tabular}{lllllll} 
\hline
Filter & $\langle \lambda \rangle$ & \multicolumn{2}{l}{Starburst}                 & Diffuse       & $L_{\rm cl}/L_{SB}$  \cr
\cline{3-4}
\noalign{\smallskip} 
& & mag & $f_\nu$ & mag  \cr
       & \AA			   & vega  & 
W/m$^{2}$/Hz & vega    &   \cr
\hline 
\noalign{\smallskip}
$x$ (F218W)$^a$ & 2210 & 12.61 & 6.6 10$^{-29}$	& 12.97     & 0.29 \cr
$u$ (F336W)$^a$ & 3348 & 13.35 & 5.7 10$^{-29}$       & 13.66     & 0.25 \cr
$b$ (F439W) & 4316 & 14.58     & 6.3 10$^{-29}$       & 14.83     & 0.21 \cr
$v$ (F555W) & 5465 & 14.08     & 8.7 10$^{-29}$       & 14.31     & 0.19 \cr
$i$ (F814W) & 8040 & 14.07     & 5.9 10$^{-29}$	& 14.37     & 0.15 \cr
\hline
\end{tabular}
\end{flushleft}
{\bf Note:} a -- corrected for UV-contamination 0.12 mag in F228W and 0.01 mag in F336W
\label{starburst}
\end{table}

\begin{figure}
\resizebox{\hsize}{!}{\includegraphics{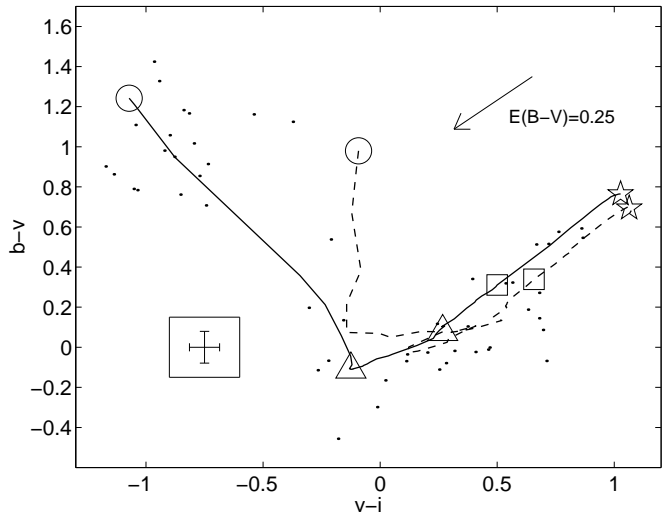}}
\caption{Colour-colour diagram ($v-i$ vs $b-v$), see caption of
Fig. \ref{xubv} for more details. The full drawn line show the 
prediction form the Z2001 model with 
Salpeter IMF, $M = 0.08$ to $120 M_\odot$, instantaneous burst, 
covering factor = 1.0, $Z_{\rm gas} = 0.002$, $Z_{\rm stars} = 0.001$. The dashed 
line shows the prediction from the publicaly available P\'EGASE.2 model 
with similar model parameters (instantaneous burst with $0.08 - 120 
M_\odot$ Salpeter IMF, $Z_{\rm gas} = 0.002$, $Z_{\rm stars} = 0.002$).}
\label{galvspeg}
\end{figure}


\section{Modelling the ages of star clusters}

The object magnitudes (corrected for aperture effects, Galactic extinction, 
and UV-contamination)
were converted into fluxes and assigned weights inversely proportional to the square
of the uncertainties. Since the uncertainties are based on the photon statistics, and there 
are other error sources present as well, e.g. zero-point and flat field errors, a 
maximum weight corresponding to an uncertainty of 0.03 magnitudes was used. 
The contribution from emission lines is significantly larger  
in the $v$ filter (F555W).  Since the nebular component is sensitive to the assumed 
ISM parameters, and because of the difficulty in isolating all the diffuse nebular 
emission associated with a particular star cluster, 
we increased the formal uncertainties with a 
factor of 2 in this filter. 
The predictions from the spectral synthesis models, having
51 time steps for the Z2001 model, is interpolated linearly onto a finer grid with an age 
resolution of 1 Myr up to 3 Gyr and 100 Myr from 3 to 15 Gyr. Finally, we find the best fitting
age for each cluster by minimising the weighted RMS deviation. When allowing $E(B-V)$ to vary, 
we used $E(B-V)=0.0, 0.05, 0.10, 0.15, 0.20$ and 0.25 and determined the best 
fitting value for each cluster.

Masses were determined by scaling the extinction corrected $v$ flux to the $M/L_v$
predicted by the model. These take into account the total mass converted into stars
and thus accounts also for stellar remnants and gas returned in the process of stellar 
evolution. The ``true'' stellar mass (excluding returned gas but including remnants) will
therefore be overestimated at high ages, but since we are  interested in 
the star formation history, we prefer using the total mass initially present in
each cluster.  At 15 Gyr the overestimate in $M/L_v$ will amount to a factor
of $\sim 1.5$ for $0.08-120 M_\odot$ Salpeter IMFs (for any $Z_\mathrm{stars}$), $\sim 3$
for $\alpha=1.35$, and $\sim 1.1$ for $\alpha=2.85$.

In Fig. \ref{galvspeg} we show the $v-i$ vs $b-v$ diagram  for clusters
with errors smaller than 0.15 mag in both colours together with a predictions from the 
Z2001 and P\'EGASE.2 codes for a standard set of parameters.
Only Z2001 is successful in reproducing the youngest clusters,
emphasizing the need for a careful treatment of the nebular contribution. For ages 
larger than a few times 10 Myr, both models approximately perform equally well.
However, around a few 10 Myrs 
the P\'EGASE.2 model presents loops in the colour-colour diagram (Fig. \ref{galvspeg})
which the Z2001 model does not. This different behaviour is due to the use of
different evolutionary tracks (see  Charlot et al. \cite{charlotetal} for a discussion). 

\subsection{Internal reddening} 

The effects of age and reddening are in general difficult
to disentangle, although the reddening and aging vectors are not absolutely
parallel. In particular, the reddening vector for young objects is close to orthogonal to the age--metallicity sequence
(Fig. \ref{galvspeg}) and, as has also been shown by other investigators (e.g. Mas-Hesse 
and Kunth, 1999), the addition of 2000\AA \ data helps in deriving the extinction 
for very young objects, since the UV-spectral 
slope is nearly constant at low ages (see Fig. \ref{xubv}).

In ESO338-IG04 we have several reasons to suspect that the internal reddening 
is very small. Groundbased  studies of the Balmer emission
line decrement (Bergvall 1986, Iye et al. 1987, Calzetti et al. \cite{calzetti}, 
Raimann et al. \cite{raimann}), as well as studies of the UV continuum (Calzetti 
et al. \cite{calzetti}, Meurer et al. \cite{meurer}, Buat et al. \cite{buat})
of the central star forming region  have given values of 
$E(B-V)$ in the range from 0 to 0.05. Further support from this conclusion comes
from a comparison of the far IR to UV luminosity (Meurer et al. \cite{meurer99}, 
Buat et al. \cite{buat}) and the
colours of the bluest star clusters themselves (this paper).  Hence, we 
know that the average extinction is very small, making ESO338-IG04 an ideal target 
to study the formation and evolution of star clusters. 
However, star formation is normally associated
with dusty molecular clouds, and it is likely that the reddening varies from cluster
to cluster. Especially we may suspect the existence of embedded very young
sources, that, however, will have quite low probability of being picked
out from this optically selected sample. Groundbased long-slit spectra, 
obtained with the ESO NTT shows a 
${\rm H}\alpha/{\rm H}\beta$ value that varies along the slit, corresponding to 
a varying  $E(B-V)$ in the range 0.0 to 0.20, although the luminosity weighted
average is close to zero. To adress the internal extinction we have
adoped two approaches: i) Allowing $E(B-V)$ to be a free parameter in the fit, but
constraining it to the range $E(B-V) \in [0,0.25]$. ii) Adopting a single extinction
for all clusters in the range $E(B-V) \in [0,0.25]$. 

In general we use a Galactic extinction law (Seaton \cite{seaton}), which has an 
intermediate behaviour in the UV compared to the LMC and Calzetti
laws commonly used for starbursts  (Calzetti \cite{calzetti}). In the 
optical, the difference is  small. Tests perfomed with the Calzetti law give close
to identical results regarding age and best fitting $E(B-V)$. Hence, the results 
presented are not sensitive to the choice of extinction law within the range of low 
extinctions considered. However, for   modelling the UV emission of clusters in
more reddened galaxies, the extinction law will be an additional source of uncertainty. 

\subsection{Comments on the parameter choices} 

In most model runs, we have kept the parameter choice fixed, but in some cases,
we have allowed some parameters to be free, as e.g. the covering factor or
the IMF slope.  There is accumulating
evidence for an IMF slope very close to the Salpeter value for stars more
massive than $1 M_\odot$ (Massey \& Hunter \cite{mh}, Kroupa \cite{kroupa},
Kennicutt \cite{kennicutt}, Leitherer \cite{leitherer}). At lower masses, the 
mass function seems to turn-over
with the result that the IMF can be approximated by a multi-segment power-law
(e.g. Miller-Scalo \cite{millerscalo}, Scalo 1998). Since the main sequence life 
time of a $1 M_\odot$ star is on the order of 10 Gyr, the shape of the lower
mass IMF will have very little influence on colours for any age considered in
this paper. The main effect lies in a different mass normalisation leading to
lower effective $M/L$ values for the Scalo98 (Scalo, \cite{Scalo}) IMF for ages 
greater than $\sim$ 300 Myr,
compared  to a Salpeter IMF.

The choice of IMF slope ($\alpha$), upper mass limit and covering factor has 
important effects on the colours at low ages. At high ages, the most important effect 
is on the mass 
to light ($M/L$) values, introducing uncertainties in the masses estimated from photometry. 

\subsubsection{Covering Factor}

In our Z2001 model runs we have used two different gas covering factors in the
calculation of the gas spectrum: 1.0 and 0.5, describing what fraction of ionising 
photons along the line of sight that will be absorbed.
This influences the strength 
of emission lines relative to the continuum, and hence the predicted flux, notably in 
 the $v$  filter. A covering factor of 0, corresponding to no
nebular emission contribution at all, was also considered, but could not fit the data
for young/blue objects. The choise of 0.5 is somewhat arbitrary, but was included
to have an intermediate value to get an estimate of the importance of this parameter.

Figure \ref{image} shows that a fraction of the ionised gas emission
is found outside the bright clusters, but this is expected since even in
the UV, young clusters don't produce more than 30--40\% of the
starburst luminosity. The ionising continuum from a young cluster will
produce significant nebular emission, that may
have a different spatial distribution compared to the clusters.  
Narrow-band  [\ion{O}{iii}] images 
 obtained with HST/STIS (\"Ostlin et al.
in prep.) show that the peaks in the [\ion{O}{iii}] surface brightness indeed
coincide with blue UV-luminous sources. However, the [\ion{O}{iii}] is more
diffuse than the stellar continuum, and our  apertures could miss a fraction
of the gas emission. Due to the ionisation structure, a larger fraction
of the [\ion{O}{iii}] emitting zone will be included as compared to the [\ion{O}{ii}] and 
H{\sc ii} zones. Since our apparently youngest clusters still have
as blue $(v-i)$ and red $(b-v)$ colours (where the $v$ band excess mainly
comes from [\ion{O}{iii}] line emission) as the earliest model steps,
the net  contribution from gas outside the aperture must be small.

Maiz-Appellaniz (\cite{maiz}) studied the spatial structure of gas emission in a sample 
of young star clusters, and found that $\sim$2 Myr old clusters have small ($\sim$10 pc) 
shells around them, whereas $\sim$4 Myr clusters have larger shells,
and for older clusters only diffuse H$\alpha$ emission is detectable. 
This indicates that with  photometric aperture diameters of $\sim$50 pc, 
we would not miss any emission 
during the first few Myr, and the covering factor should be close to unity. 
At later times 
(5-10 Myr), we may miss a significant fraction of the emission, which can 
be approximated by a smaller covering factor. 
At still larger ages the clusters will be too cool to ionise [\ion{O}{iii}] and
the ionisation structure will be of little concern. Moreover, the relative
gas contribution will be almost negligible in all passbands used here.


\subsubsection{Metallicity}

The nebular oxygen abundance in ESO338-IG04, as determined using the 
temperature sensitive [O{\sc iii}]$_{\lambda 4363}$  line, has been shown to be on the 
order of 10-14\% solar (Bergvall 1984, and Masegosa et al. 1994). Hence, 
in general, we adopt $Z_{\mathrm{gas}}=0.002$, ($Z_\odot=0.0188$).
Other values (see Table \ref{z2001}) are also 
included to test how the assumed nebular abundances affect the results. It 
is comforting that the best fit is provided by $Z_{\mathrm{gas}}=0.002$.

For the stars, we have very limited information on the metallicity. In
general Fe, which is the dominant source of stellar opacity, is 
less abundant (compared to solar) than O in metal-poor galaxies (Kunth and \"Ostlin 2000).
Also, in a closed box model, the stars should on average have lower
metallicity than the gas. Hence, the choice $Z_{\mathrm{stars}}=0.001$
seems reasonable. However, there are indications from near-IR surface
photometry that the underlying  old host 
galaxy may be more metal rich than the nebular gas (Bergvall and \"Ostlin 2002).
This intriguing result can be interpreted as metal-poor gas being accreted 
onto a more metal-rich host galaxy in a merger.
Whereas young objects should have a metallicity  close 
to that of the gas, it cannot be excluded that older clusters are
significantly more metal-rich. However, metal-rich elliptical
galaxies have in general metal-poor (typically $[Fe/H]<-1$, Brodie and Huchra \cite{brodie})
old globular cluster systems. The commonly found metal-rich sub-populations
are in general younger and still less metal-rich than the host galaxy
(Ashman \& Zepf \cite{ashmanzepf}).

\subsubsection{Star formation time-scale}

The assumed time-scale for star formation $\tau_{\rm SF}$, i.e. how long
star formation is active in a particular event, impacts the 
model predictions for ages comparable to  $\tau_{\rm SF}$. As
long as star formation is active, young stars will make colours 
blue and modelled ages tend to increase with the assumed $\tau_{\rm SF}$.
This is a major source of uncertainty when modelling the integrated light 
from galaxies.  In these compact ($r \sim$ a few pc) star clusters we 
know that $\tau_{\rm SF}$ must be short, which is a main reason why ages 
of such are less degenerate than those for unresolved populations.
Of course, the common assumption of a truly instantaneous formation of a massive star cluster
is unrealistic. 
Another question is whether low- and high-mass
stars form simultaneously, or with some time lag inbetween. When massive stars  
appear, they ionise their surroundings and inhibit further star formation. 
Hence,  low mass stars  must form
prior to or simultaneously with the massive stars. McKee and Tan (2002) argues 
that the formation time scale for individual stars is short, of the order of 
$10^5$ years, permitting the coeval formation of low and high mass stars in 
a cluster.  Massey and Hunter (1998) found that the formation of R136, the 
central cluster in 30 Doradus, might have been extended over $\sim$4 Myrs, with 
low mass stars forming prior to  massive ones. Moreover the IMF slope 
is $\alpha=2.3 - 2.4$ over the studied range 2.8-120 $M_\odot$, i.e. identical
to the Salpeter (1955) value. Similarly, the central star cluster in
the Galactic giant H{\sc ii} region NGC3603 shows an age spread of a few Myr
(Pandey et al. 2000).

Our three main SFHs considered are instantaneous burst, a constant SFR for
10 Myr after which the SFR drops to zero, and an exponentially declining
SFR with an e-folding time of 3 Myr. A smaller number of other parameter options
 were tested 
as well (Table \ref{z2001}).  A potential drawback with the exponential
 model is that the SFR only goes 
asymptotically to zero, and hence remains significant for quite a long time.
Although little of what is derived in this paper will be substantially 
changed whether one adopts either of these three timescales,
we will show below that comparison with the global H$\alpha$ output tend
to favour the shorter one. An instantaneous burst also gives
on average lower RMS residuals. 

In all our evolutionary sequences, low- and high-mass stars form simultaneously. If in reality 
there is  a time lag of a few Myr in the formation of the most massive 
stars, this will influence our results for ages on the order the time lag. 
For a time lag of 3 Myr our model 
would behave as follows: Looking at Fig. \ref{imftest} for reference, 
the time-lag tracks would start near $v-i=-0.1, b-v=0$ and climb along the 
simultaneous tracks up to $v-i=-1.2, b-v=1.2$, a point it would reach when the 
most massive stars appear after 3 Myr. Then the spectrum would essentially 
follow the same track as for the simultaneous case model. Hence the only
effect would be that we would confuse ages during the first 3 Myr. We will
discuss the possible implications of time-lag on our results when relevant.

\begin{figure}
\resizebox{\hsize}{!}{\includegraphics{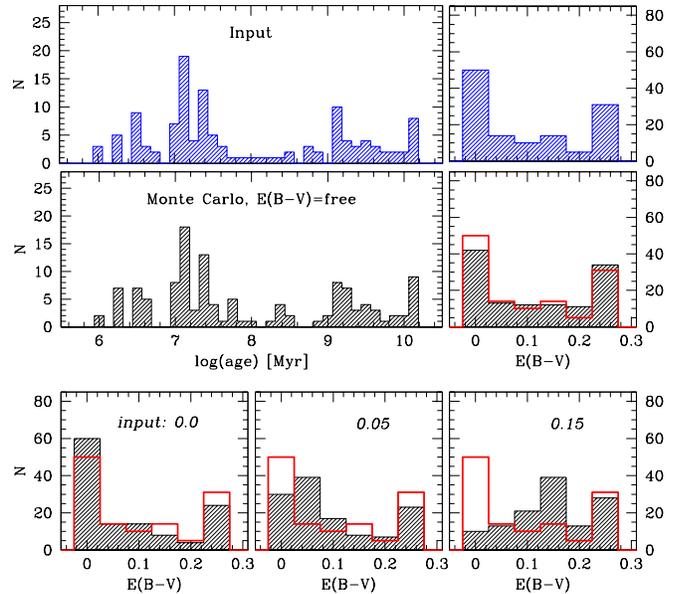}}
\caption{Some results from the Monte-Carlo simulations.  
{\bf Top row:} Results from a typical evolutionary sequence (Salpeter IMF, $M = 0.08$ to $120 
M_\odot$, 10 Myr burst with constant SFR, covering factor 1.0, $Z_{\rm gas} = 0.002$, 
$Z_{\rm stars} = 0.001$). On the left we show the age distribution, and on the 
right the   $E(B-V)$ distribution.
This age and $E(B-V)$ distribution was used to create a fake catalogue 
where the objects match the evolutionary sequence perfectly in all filters.
To this fake catalogue, random errors were added  according to
the observed uncertainties, and 
then used as input to the fitting program. 
{\bf Second row:}  Sample output from this experiment 
when $E(B-V)$ was a free parameter. In the $E(B-V)$ histogram, 
the ``input'' $E(B-V)$ distribution of the
top panel is shown as a thick unshaded histogram. 
{\bf Third row:} Output $E(B-V)$ distributions from
sample Monte-Carlo simulations where the input fake catalogue was created
with fixed values of $E(B-V)$. The input $E(B-V)$ value is given 
as a number in each sub-panel, and the  $E(B-V)$ distribution of the top panel 
is shown as a thick unshaded histogram for reference.
}
\label{fakehists}
\end{figure}

\subsection{Monte-Carlo simulations} 

In order to assess the internal accuracy of the fitted ages and extinctions,
we constructed  catalogues of artificial clusters with the same age and luminosity 
distribution as indicated by the model fits and with colours that perfectly fit 
the models. To these ideal clusters we added observational errors in accordance 
with the observed error distribution. Then this fake catalogue
was fitted to the models. The experiment was repeated  100 times to
beat down the statistical noise and from the dispersion of the fitted ages we
could for each object assess the uncertainty in the derived age, for the
assumed parameter configuration in question. The top two panels in Fig. \ref{fakehists} show
the input (top) and output (second row) age and $E(B-V)$ distributions for
a sample Monte-Carlo experiment. The tendency to slightly overestimate the
output extinction  is statistically significant, but
is only a few hundredths of a magnitude on average. 

We also created fake catalogues with fixed input $E(B-V)$ values.  
Specifically, we wanted to test whether the $E(B-V)$ 
range found from the free fit could be an artifact produced from an
(in reality) single  reddening 
value smeared out by observational errors. The bottom panel in Fig. \ref{fakehists}
compares the reddening found from the free fit method to simulations with 
fixed input $E(B-V)$.  Obviously, no single reddening value can be 
found which can reproduce the distribution 
found when letting $E(B-V)$ be a free parameter, although $E(B-V)=0$ comes quite
close. Hence, we conclude that there is a real spread in $E(B-V)$, but  
letting it be a free parameter on average may lead to a  slight overestimate of 
the true $E(B-V)$. 

Since the colour and luminosity of a cluster changes with time, 
the  uncertainty in the derived ages is not time-independent.  We have  used the Monte-Carlo simulations to estimate the internal 
uncertainties on the derived ages resulting from observational errors.  The net effect is 
small uncertainties ($\sigma_{\rm Age} \approx 0.1$ dex) for ages less than 100 Myr, 
considerable uncertainties in the range 100 Myr to 1 Gyr ($\approx  0.4$ dex), 
and again quite small uncertainties for ages greater than 1 Gyr ($\approx 0.2$ dex).
These numbers are similar for any
evolutionary sequence with $Z_{\mathrm{stars}} \le 0.004$. 
For higher metallicities the age of many objects decreases, 
and the region of higher uncertainties at intermediate ages is shifted
to smaller ages.
Our conlusion is that  the overall age pattern is quite 
insensitive to observational errors.


\begin{figure*}
\resizebox{\hsize}{!}{\includegraphics{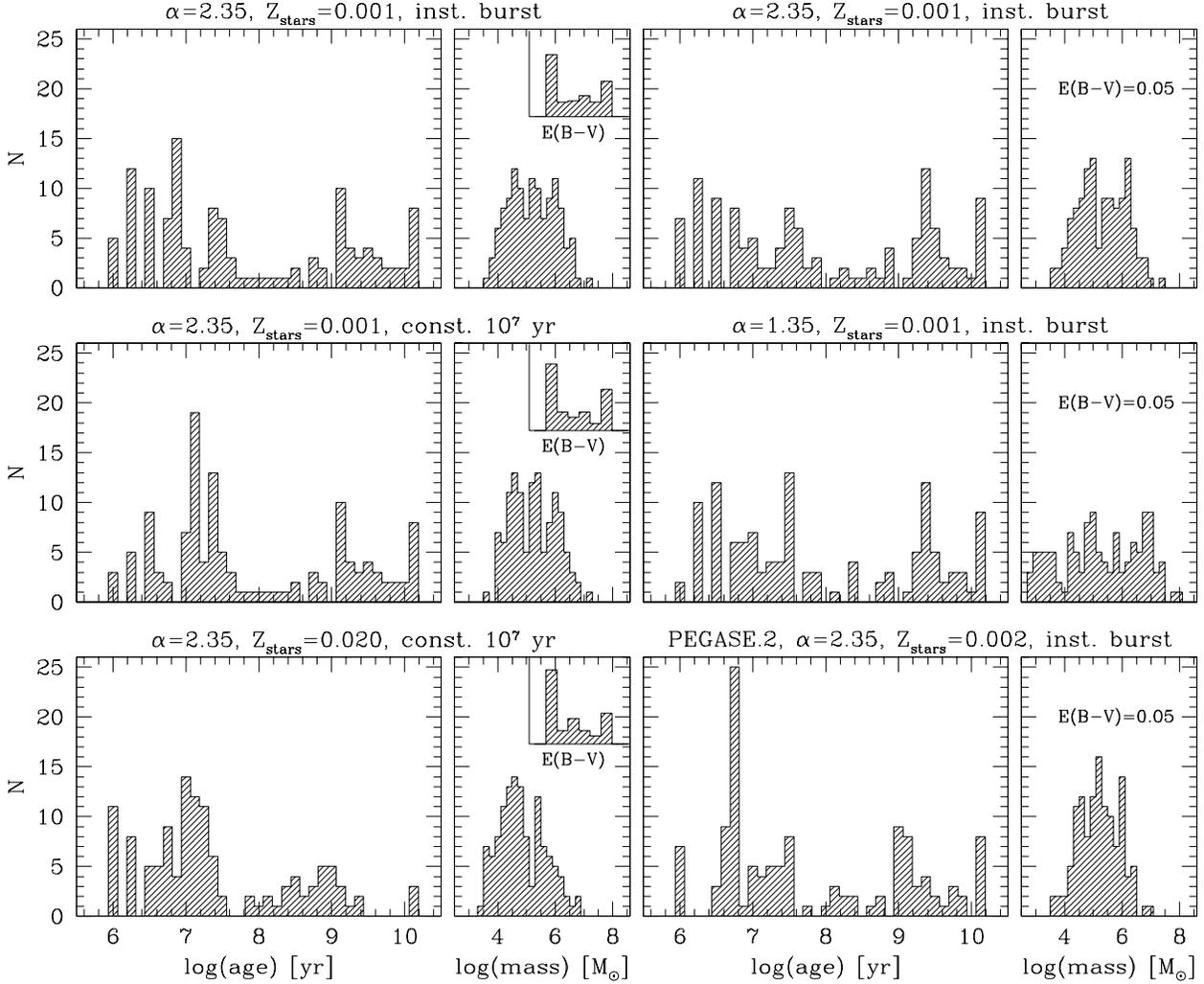}}
\caption{Comparison of output age and mass distributions for different model assumptions.
In the left panels, the extinction was allowed to vary freely in the range $E(B-V)\in [0,0.25]$,
and the resulting exinction distribution is indicated by the inset in the mass histograms.
In the right panels a fixed extinction of  $E(B-V)=0.05$ was used.
All evolutionary sequences, except for the lower right one, was calculated with Z2001 code, and have the 
following parameteers in common: Mass range $M = 0.08$ to $120 M_\odot$, 
$Z_{\rm gas}=0.002$, and covering factor 1.0. Other parameters:
{\bf top left:} Salpeter IMF ($\alpha=2.35$), $Z_{\rm stars}=0.001$ and instantaneous burst.
{\bf middle left:} $\alpha=2.35$, $Z_{\rm stars}=0.001$ and constant SFR for 10 Myr.
{\bf lower left row:} $\alpha=2.35$, $Z_{\rm stars}=0.020$ and constant SFR for 10 Myr.
{\bf top right:} Salpeter IMF ($\alpha=2.35$), $Z_{\rm stars}=0.001$ and instantaneous burst.
{\bf middle right:} $\alpha=1.35$, $Z_{\rm stars}=0.001$ and instantaneous burst.
{\bf lower right:} PEGASE.2 model $\alpha=2.35$, $Z_{\rm stars}=0.002$ and instantaneous burst.
}
\label{agehists}
\end{figure*}

\section{Results}

We find an overall temporal age pattern that is quite robust to changes in all 
model parameters
except $Z_{\mathrm{stars}}$ and the lower mass limit. The latter matter only for
ages comparable to the lifetime of stars with mass equal to $M_{\rm low}$,
and $M_{\rm low} = 2 M_\odot$ is  unrealistic for 
ages $>1$ Gyr. Despite some claims of  IMFs devoid of 
low mass stars in starburst regions, there is no direct evidence for such, and
we mainly consider such evolutionary sequences for comparison.

In Fig. \ref{agehists} we show age and mass histograms for a selection of evolutionary sequences 
with $E(B-V)$ as a free parameter and fixed $E(B-V)=0.05$. 
In accordance with \"OBR98, there is evidence both for young (less than 100 Myr)
intermediate (a few Gyr) and old (more than 10 Gyr) clusters if a low metallicity
is adopted. If a high metallicity is adopted, the net effect is to compress
the age span, but retaining the main features. A metal-rich stellar component 
provides a poor fit to the bluest objects which  must therefore have
a metallicity close to that derived for the nebular gas. 
Adopting a nebular metallicity of $Z_{\rm gas}=0.002$, provides for a significantly better fit for young clusters,
which give us confidence in our way of handling the nebular emission in Z2001.

The shape of the mass histograms are model dependent (due to
the different mass normalisations inherent to each IMF), but for most parameter choices we find
that the maximum mass is higher for clusters older than 1 Gyr, and more 
mass seems to be found in relatively old than  young clusters. All evolutionary sequences give 
a mass function (MF) with many clusters with $M>
10^6 M_\odot$, showing that this galaxy is rich in massive globular clusters.

Comparing the free fit $E(B-V)$ with the $E(B-V)=0.05$ results  (top two panels in
Fig. \ref{agehists}),  
one sees that peaks in the age distribution tend to be displaced to 
lower ages when $E(B-V)$ is allowed to vary freely. This
effect is notable for the peak at an age of a few Gyr. The peak is there for all
metal-poor evolutionary sequences but occur at 0.2 dex lower age for the free fit $E(B-V)$ cases.
At these ages, few sources have accurate photometry in more than 2 bands ($v$ 
and $i$) and the fixed $E(B-V)$ results are probably more robust and less prone 
to systematic errors. However, the overall results in terms of age and mass 
distributions are insensitive to the adopted extinction treatment, but this is 
so only because we have a priori information that the extinction must be low.

When adopting a fixed $E(B-V)=0.05$ (as in \"OBR98) we find a good overall  
agreement with the result in \"OBR98, except for ages below 
10 Myr where the evolutionary sequences used in \"OBR98 had no 
time resolution.  We do not find the peaks at 100 and 600 Myr reported by  
\"OBR98, although we do find objects with these ages.

The age pattern is sensitive to the adopted star formation time-scale $\tau_{SF}$
at ages comparable to  $\tau_{SF}$ (compare top-left and mid-left panels in 
 Fig. \ref{agehists}). The IMF influences the ages, 
but the effect is  small at high ages; the effect on masses is more
severe (upper and middle right panels in  Fig. \ref{agehists}).  
Both the $\alpha = 1.35$ and 2.85 evolutionary sequences have problems at high ages since
the predicted mass to light ratios (for 15 Gyr: $M/L_V = 47$ and 8.7, respectively, as measured 
by the total consumed mass, whereas $\alpha = 2.35$ give $M/L_V = 6$\footnote{A Scalo98 
IMF would give a factor of $\sim$2 lower$M/L_V$} or $M/L_V = 4.2$ if counting 
only stars and remnants) is much higher than those 
observed in old Galactic globular clusters ($M/L_V\sim 2$, Ashman \& Zepf \cite{ashmanzepf}).
Gas parameters are only significant for ages smaller than or comparable to a few
times $\tau_{SF}$. The found $E(B-V)$ 
distribution is relatively insensitive to the adopted parameters.
Within each parameter configuration we have analysed the correlation between the fitted age, mass,
residual and $E(B-V)$.  Mass and age are corellated (as expected since  $M/L$ is
a strong function of age), but not the other quantities.

We have performed 
correlation analysis on the output from the different parameter configurations. All evolutionary sequencecs with 
a low stellar metallicity and $M_{\rm low}=0.08 M_\odot$ show a high degree 
of mutual correlation, corroborating the results in Fig. \ref{agehists}.
The correlation with more metal rich sequences is significantly weaker but these 
parameter configurations viewed as a group show a high degree of internal correlation. 
Masses, RMS residuals and $E(B-V)$ values between different model runs are tightly 
correlated whenever the age-correlation is high.

\subsection{Comparison between Z2001 and P\'EGASE.2}

Figure \ref{agehists} also show the results from one of the P\'EGASE.2 parameter configurations.
 We have already noted
that the different treatments of the nebular component result in very
different behaviour  during the first 10 Myr, and 
that P\'EGASE.2 cannot account for the most extreme colours. 
Also at later ages there are some differences between the two
models. For a fixed $E(B-V)$ the qualitative agreement for ages larger 
than 50 Myr is fair (Figure \ref{agehists}), although  P\'EGASE.2 give 
lower ages on average.
For the $E(B-V)$ free  parameter cases, the pronounced peaks at 
$\sim1$ and 15 Gyr are replaced by a broader distribution for the P\'EGASE.2 
model. Comparing the tracks in Fig. \ref{galvspeg} and 
\ref{xubv} it is not surprising that we get different results, as
 the P\'EGASE.2 model show loops in the $bvi$ diagram (Fig. 
\ref{galvspeg}) between 10 and 500 Myr and a redder colour at 1 Gyr.
For both Z2001 and P\'EGASE.2, a potential danger with a free fit $E(B-V)$ is that 
features in the evolutionary sequences, e.g. loops and kinks, may act as attractors.
The degree of correlation between results from Z2001 and P\'EGASE.2 
with similar parameter settings is high for age and mass but 
 smaller  for $E(B-V)$ and RMS residuals.

\section{Resulting constraints on the starburst physics}

In agreement with the spectroscopic analysis, we find that young objects
(age $<$ 20 Myr) are better reproduced by metal-poor ($Z_{\rm stars} \le 0.004$) 
stellar components. At somewhat higher ages,
 a slightly higher stellar metallicity ($Z_{\rm stars}=0.008$) 
is also allowed and in some cases gives a slightly better fit.
For ages on the order of 100 Myr and above, results are inconclusive
regarding metallicity.
However notable, if a higher metallicity
is adopted, much fewer objects with ages in excess of 1 Gyr are found.

\begin{figure}
\resizebox{\hsize}{!}{\includegraphics{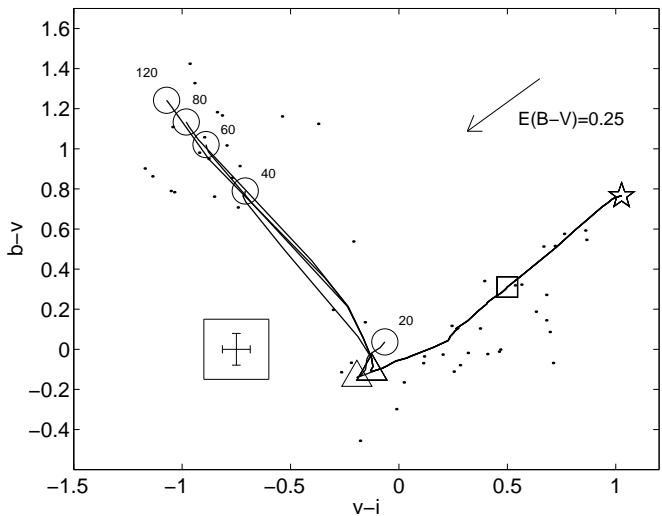}}
\caption{Colour-colour diagram ($v-i$ vs $b-v$) showing, as dots, 
the location of clusters with uncertainties smaller than or equal 
to the rectangle in the lower left. The median uncertainty is 
indicated by the cross. The arrow indicates how  objects would be 
repositioned after correction for a hypothetical extinction of 
$E(B-V)=0.25$. Lines represent the evolutionary
tracks of the models from 0.5 Myr (circles), 10 Myr (triangle), 
1 Gyr (square) to 15 Gyr (Star). The predictions at the youngest 
age are labelled with the assumed upper mass limit. Parameters
common for all scenarios: instantaneous burst,  covering factor = 1.0, 
$\alpha=2.35$, $M_{\rm low}=0.08 
M_\odot$, $Z_{\rm gas} = 0.002$, $Z_{\rm stars} = 0.001$.}
\label{imftest}
\end{figure}

We next turn our attention to the  upper mass limit and slope of the IMF, which is
important for regulating the ionising flux during the early stages 
of a burst. Hence it influences the UV-continuum and the ionised
gas spectrum. The combination of red $b-v$, and blue $v-i$  colours in
the colour-colour diagram  at low ages in many evolutionary scenarios is a direct 
consequence of significant contribution from gas emission to
the $v$-filter (F555W) whereas the $b$ and $i$ filters (F439W and F814W) 
are free of strong lines, although nebular continuum is present
in all filters.  
All scenarios with an upper mass limit of $M_{\rm up} = 20 M_\odot$ fail 
to reproduce the objects with red $b-v$ and blue $v-i$ colours (Fig. \ref{imftest}). 
Can we try to pin down the minimum allowed upper mass limit? A necessary requirement
must be that such a scenario can reproduce the cloud of objects in the upper 
left part of Fig. \ref{imftest}. To investigate this issue, we ran a new
set of evolutionary sequences with $M_{\rm up} = 40$, 60, and $80 M_\odot$ using Z001.

Let us first look at the Salpeter IMF. For a covering factor of 0.5 
 the model cannot reproduce the observations even if  $M_{\rm up} 
= 120 M_\odot$. For a covering 
factor of 1.0, there is no problem and
an upper mass limit of $80 M_\odot$ can be 
accepted (Fig. \ref{imftest}). However, if we increase 
the filling factor,  we can mimic the effect of increasing $M_{\rm up}$. 
Hence, even if a filling factor of unity
appears  unrealistic according to the current knowledge of the ISM (Kassim et al. 
\cite{kassim}), we cannot rule
out that it is higher than our generally assumed value of 0.1.
Our  more conservative conclusion is therefore that for a Salpeter IMF, 
$M_{\rm up} \ge 60 M_\odot$. This is in agreement with Bergvall (1985) 
who from the observed [O{\sc iii}]$_{\lambda5007}/{\rm H}\beta$ 
ratio found that $M_{\rm up}> 50 M_\odot$.

\begin{figure}
\resizebox{\hsize}{!}{\includegraphics{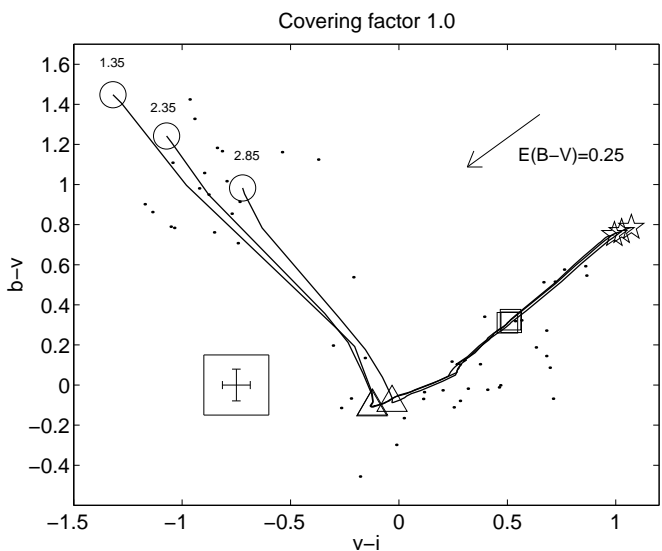}}
\caption{Colour-colour diagram ($v-i$ vs $b-v$) showing the location 
of clusters and evolutionary scenarios with different IMF slope. Symbols have the
same meaning as in Fig. \ref{imftest}. The scenarios are labelled with 
their assumed IMF slope: $\alpha = 1.35$, 2.35 and 2.85.
Other parameters common for all scenarios: instantaneous burst 
$M_{\rm low}=0.08 M_\odot$, $M_{\rm up}=120 M_\odot$, $Z_{\rm gas} = 0.002$, 
$Z_{\rm stars} = 0.001$, covering factor $= 1.0$.}
\label{imfslopetest}
\end{figure}

For  $\alpha=1.35$,  a covering factor of 1.0 give very extreme colours
(Fig. \ref{imfslopetest}), suggesting that cluster formation has already ceased 
or that $M_{\rm up} \le 60 M_\odot$. Even  a covering factor of 0.5 give quite extreme 
colours at the earliest times, but in general offset (to bluer $v-i$ colours) 
from the cloud of points. This offset could possibly
be explained by an average extinction of $E(B-V)=0.15$, which is much more than 
the estimates presented in Sect 4.1. 
 If the sources are allowed to have 
a range of covering factors, an upper mass limit as low as $M_{\rm up} \ge 40 
M_\odot$ would  provide a decent fit and the extinction may be lower.
If we adopt $\alpha=1.35$ we must assume 
that either: {\boldmath $i$}) there are plenty of objects with age 10 Myr, but very few ones 
younger than this, or {\boldmath $ii$}) $M_{\rm up} \le 60 M_\odot$,
or {\boldmath $iii$}) the covering factor is 0.5 and the extinction is higher 
than previous estimates.

For $\alpha=2.85$, a covering factor of 1.0 in combination with 
$M_{\rm up}=120 M_\odot$ is needed to at least touch upon the  
the cloud of young clusters (Fig. \ref{imfslopetest}). Hence we would 
be forced to adopt the
lowest possible ages together with uniform high filling and covering
factors. We regard this as unrealistic, and hence $\alpha=2.85$ is not
our preferred IMF.

Could the IMF vary from cluster to cluster? The issue whether the IMF is 
universal or not has been subject to a many discussions (see Kroupa \cite{kroupa}). 
Variations could either be due to environmental or 
stochastical effects. The environmental effects should be similar for
all clusters in ESO\,338-04, but in low mass clusters the upper mass
range of the IMF may be badly populated and subject to statistical noise.
For a Salpeter IMF with 0.08--120  $M_\odot$, the optical/UV emission from
young  clusters may be affected by stochastical effects for cluster masses 
below $10^5$ $M_\odot$ (Lan\c{c}on and Mouhcine \cite{lancon}),
which is close to our median mass for your young clusters. 
Hence we cannot exclude stochastical effects 
among  the fainter  young clusters which could mimic a 
variable IMF slope. But in this case the average must be close to
$\alpha= 2.35$.

\section{Clustered vs diffuse emission}

At first sight (Fig. \ref{image}, \ref{image2}) it appears that the star burst region is
dominated by emission from compact star clusters. However from Table 3
 it is clear that even the central 1 kpc is
dominated by diffuse emission. Of course,
the diffuse emission is also likely to be clustered at some level and
include open clusters, OB associations etc.

The integrated luminosity distribution of clusters in ESO\,338-04 appears
to have a Gaussian shape  with a peak at $M_V\approx-10$ (\"OBR98) but this is
primarily a consequence of the photometric completeness limit and a mix of 
ages with different $M/L$. In each age
bin containing a significant number of clusters, the luminosity function (LF)
is best described by a power law $dN/dL \propto L^{-\beta}$, with power
law index $\beta = 1.8 \pm 0.15$ down to the completeness limit. This 
is valid also for the F218W data. From this we can estimate, roughly, how
large a fraction of the clusters that fall below the completeness limit.
We estimate, that if we correct for clusters below the detection threshold,
the clusters would make up 40\% of the light in F218W. However, although
the number of young clusters is small outside the central kpc, diffuse
H$\alpha$ emission is present out to much greater radius (Bergvall \& 
\"Ostlin 2002). Is this gas ionised in situ by a more diffuse population
of young stars, or by leakage of Lyman continuum photons from the central
burst? The H$\alpha$ emission inside $r=1.075$ kpc corresponds to 80\%
of the total H$\alpha$ emission in ESO\,338-04, which may bring the fraction
of the total UV emission from the galaxy associated with young clusters, back 
down close to 30 \% if
the ionisation is in situ. Hence, the fraction of  2000\AA \ emission, and 
hence young stars, coming from young clusters should be in the range 30 to 
40\%. 

Meurer et al. (\cite{meurer}) found  only 16\% of the  2200
\AA \ emission in ESO\,338-IG04 to come from clusters, but this 
was estimated in a different way (using  de-convolution 
of the pre-refurbishment HST/FOC images). The average value for nine starbursts 
in Meurer et al. (\cite{meurer}) is 20\%. We have simply integrated the
emission coming from clusters  (including a significant aperture correction) 
in all passbands and compared to the integrated emission over a circularly 
defined starburst area. Also our integrated starburst magnitude (see Table 
\ref{starburst}) is brighter by $\sim$1 magnitude compared to Meurer et al.
(\cite{meurer}). Apparently, our WFPC2 observations have been more sensitive
and the fact that we get similar numbers for the other passbands, and find
a monotonic increase in the cluster fraction with shorter wavelength, give 
us full confidence in our numbers. 

Zepf et al. (\cite{Zepf et al.}) found that star clusters are responsible
for 19\% of the blue light in the  the gas-rich merger \object{NGC 3256}, 
similar to our value of 21\%, but most other mergers have 
a smaller fraction of their optical light coming from young clusters (\"Ostlin
\cite{ostlin2000}), e.g. in the Antennae it is less than $10$\% (Whitmore \& Zhang
\cite{whitmore&zhang}).

\subsection{The ionisation energy budget}

>From the Z2001 evolutionary scenarios, we have calculated for each cluster, the
predicted output of Lyman continuum photons, and hence the predicted total
H$\alpha$ output associated with  young clusters. 
The values we get have been scaled up with a factor $\sim 1.3$ to account 
for clusters fainter than the detection threshold. This correction has been 
applied to  the estimates presented
below.

The total observed H$\alpha$ flux from ESO\,338-04 is $\log({\rm H}\alpha)=41.7\pm 0.2$ 
(\"Ostlin et al. 1999) which has only been  corrected for Galactic extinction, but since 
the internal extinction is estimated to be very small, $E(B-V) \le 0.05$, it will 
serve our purpose as an estimate. If the clusters make up 30 to 40\% of the present star 
formation, they should then produce $\log({\rm H}\alpha)=41.2$ to $41.3 (\pm 0.2$). 

A standard Salpeter metal-poor scenario with instant burst  produces 
$\log({\rm H}\alpha)=41.1$. Hence some 
emission is lacking, but it is within the uncertainty. A constant
SFR for 10 Myr gives 42.1, i.e. much larger than the 
observed value. With a constant SFR for 5 Myr  we get 41.9 and
with an exponentially decaying SFR (e-folding time 3 Myr) we get 41.6. 
An IMF with $\alpha = 1.35$ give  similar values to $\alpha = 2.35$.
Hence these results support that the star formation time-scale is short,
a few Myr at most.

\section{Star formation history as derived from clusters}

The formation of a massive star cluster is by necessity associated with
star formation in the galaxy. In the present starburst in ESO\,338-04,
30 to 40\% of the newly formed stars are found in clusters.
Are compact/massive star clusters formed as a natural consequence of star 
formation, or are special conditions, e.g. the high ISM turbulence 
prevailing in merging galaxies or circum-nuclear rings, required?
 In the former scenario, clusters form now and then as long as star 
formation is active. 
Star formation is normally a collective phenomenon, but only rarely 
(e.g. the Arches cluster, Serabyn et al. \cite{serabyn}) do
we find young stars to form self-gravitating systems in our Galaxy.
This could be a consequence of a low star formation rate density

The ability of young clusters to develop into globular cluster-like
systems is connected to their environment: Whether the clusters will 
be gravitationally bound  or not, depends 
on e.g. the ISM density, pressure and metallicity. Moreover, the
ability for clusters to survive disruption depends
on their environment (Meylan \& Heggie 1997).

For a given cluster age, the photometric completeness limit corresponds
to a given stellar mass, producing  a time dependent mass detection 
limit. The fraction of  clusters in each age bin
 brighter than the limiting magnitude depends  the cluster LF, a 
quantity which has been subject 
to some debate during the last years. The shape of the LF of old
GCs in our Galaxy is  Gaussian with a
peak of $\log(M/M_\odot)=5.3$ and dispersion $0.5$ (Fall \& Zhang 
\cite{fall}). On the other hand, the LF and MF of young clusters, e.g. in 
the Antennae (Whitmore et al. \cite{whitmoreetal}) and ESO\,338-04, is 
described by a power law without any sign of a turnover at a characteristic 
mass $\sim 10^5 M_\odot$. As clusters age and are subject to internal and 
external dynamical effects, the lower part of the mass spectrum will
be de-populated and the MF converted from power law to 
Gaussian (Fall and Zhang, 2001).
However, the initial MF may still be a power law.
Whether clusters survive or not, the power-law MF measures the mass initially
formed in  star clusters. 

%
%
%

If the initial cluster mass function is universal and known, it is straightforward 
to estimate what fraction of the clusters at a given age
that fall below the detection threshold. However, in practise the mass integral
for a power law with unbounded mass intervals diverges, and it is 
neccessary to fix the mass range, just like for the stellar IMF.
Unfortunately, these quantities are not observationally known. 
One way would be  to self consistently solve for the upper mass 
limit, but this only works when the MF is well populated.
Moreover, this violates  the assumption of a universal cluster MF
and the same applies if the upper mass limit is taken as the observed maximum 
mass in the age bin in question.
Our finally adopted method is quite crude and assumes a universal
power-law initial mass function with a mass range 
$ 10^3 M_\odot$ to $5\times 10^6 M_\odot$.

The time dependent
 correction factor for clusters below the detection treshold 
can now be calculated and  ranges from 1.25 at low ages, where we see most of
the mass (at least in systems with small reddening), to $\sim5$ at 15 
Gyr where we see only about 20\%. Assuming a Gaussian 
initial  cluster mass function   (e.g. to simulate that many 
low mass  clusters get disrupted) would
give similar numbers.  It is not vitally important
 if clusters are gravitationally bound or not. Our
estimate will always give a lower limit to the cluster SFR.

For each scenario, we have included the uncertainty of $\pm 0.15$ in 
the power law index $\beta$ of the LF and an
uncertainty in the completeness limit of 0.25 mag. The
completeness limit used is $\sim 1$mag brighter than the photometric
limit. This is to get closer to the real 100\% 
completeness, and also removes the most uncertain objects. 

\begin{figure}
\resizebox{\hsize}{!}{\includegraphics{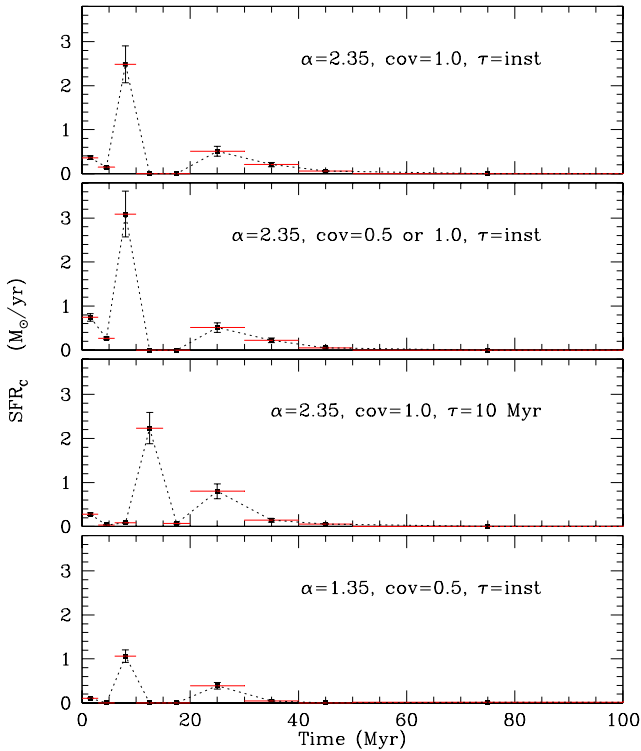}}
\caption{The star formation rate derived from clusters as a function of
time during the past 100 Myr for 4 different evolutionary scenarios,
all  with $Z_{\rm gas} = 0.002$, 
$Z_{\rm stars} = 0.001$ and mass range 0.08 -- 120 $M_\odot$.
{\bf First (top) panel}: Salpeter IMF ($\alpha=2.35$) with instantaneous burst and covering factor = 1.0. 
{\bf Second panel}: Salpeter IMF with instantaneous burst when the covering factor is allowed to vary. 
{\bf Third panel}: Salpeter IMF with a burst duration of 10 Myr and covering factor 1.0.
{\bf Fourth panel}: $\alpha=1.35$ IMF with instantaneous burst and covering factor  0.5.
Error bars in the X-direction 
shows the width of the time window, and points are
connected with dotted lines  to guide the eye.
Error bars in the Y-direction account for the uncertainty
of $\pm 0.15$ in the power law index $\beta$ of the LF and an
uncertainty in the completeness limit of 0.25 mag.}
\label{sfhshort}
\end{figure}

\subsection{The young starburst}

Figure \ref{sfhshort} shows the SFR vs time during
the last 100 Myr as derived from clusters for 4 evolutionary scenarios. The patterns 
look close to identical. For instantaneous burts the main features are: 
a small peak at $0 < t \le 3$ Myr,
followed by a decrease for $3 < t \le 6$ Myr, after which the main
peak occur in the interval $6 < t \le 10$ Myr. 
Another 
significant peak occurs at $20 < t \le 30$ after which the cluster 
SFR drops slowly but remains positive until 40 Myr.   

For  longer bursts (constant SFR during 10 Myr, 3rd panel in Fig. 
\ref{sfhshort}) the highest 
peak is shifted to the interval  $10 < t \le 15$ Myr, while the
peak at 25 Myr is not affected (see third panel in Fig. \ref{sfhshort}).
Changing the IMF does not affect the pattern, only the scaling of the 
SFR.

The relatively low SFR in the first bin ($0 < t \le 3$ Myr) is not 
a consequence of a few number of clusters, but is due to their
low estimated total mass. The scarcity 
of objects with ages near 5 Myr can be seen also in Figs. \ref{imftest} 
and \ref{galvspeg}: many objects are clustered around $v-i=0, 
b-v=0$ and $v-i=-1, b-v=1$, but few in between. Clearly, at such
small ages with rapid stellar evolution, ages will be sensitive to
details in the stellar models. The dip at 15 Myr  could 
possibly be attributed to  features in the multi-dimensional colour 
plane described by the scenarios. 

Whether or not the SFR went down to zero some 15 Myr ago, the star 
formation activity began to rise already 40-50 Myr ago, and appears 
to have peaked a little less than 10 Myr ago. Hence the starburst
in ESO338-IG04 is 40-50 Myr old and still active. This estimate is
based on clusters selected from their bright optical and UV emission.
It is possible that a fraction of the youngest sources
are hidden in dusty cocoons, that should be visible as mid-IR
sources like those found in  the BCG He2-10 (Vacca et al. \cite{vacca}),
but see discussion below.

If there is a time-lag of a few Myr in the formation of the most 
massive stars (see discussion above), some of the clusters in the 
5-10 Myr peak would instead be very young ($\sim 1$ Myr) and the
difference in SFR between the 0-5 and 5-10 Myr intervals diminished.

\subsection{The SFR viewed over a cosmological time scale}

\begin{figure}
\resizebox{\hsize}{!}{\includegraphics{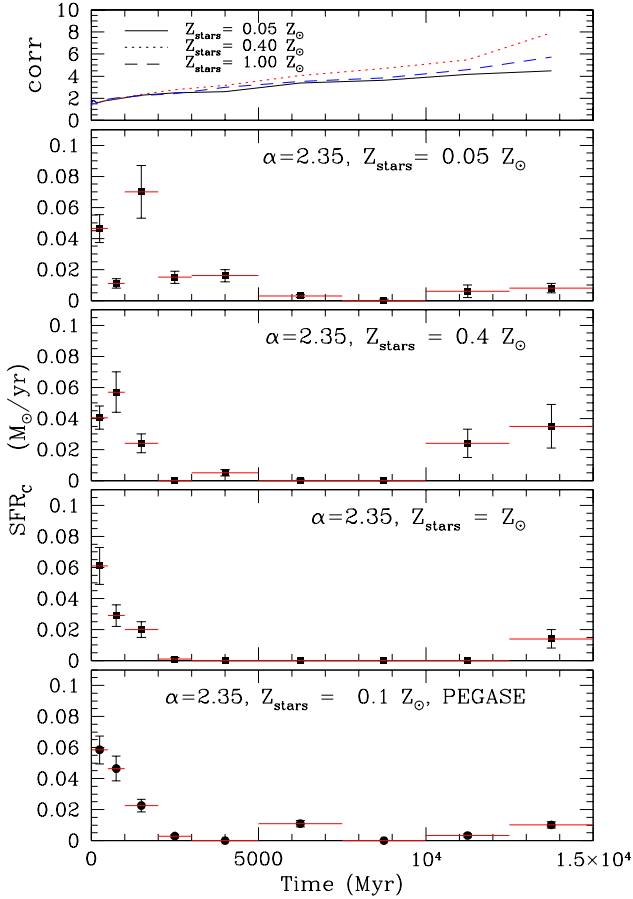}}
\caption{The star formation rate in clusters as a function of
time during the past 15 Gyr for three Z2001 evolutionary sequences with
different $Z_{\rm stars}$ (0.001, 0.008, and 0.02), and one 
P\'EGASE.2 sequence with $Z_{\rm stars} = 0.002$. 
All sequences have $\alpha=2.35$, 
mass range 0.08 -- 120 $M_\odot$ and $Z_{\rm gas} = 0.002$.
The top panel shows the correction factor due to clusters
below the detection limit for the different metallicities.}
\label{sfhlong}
\end{figure}

Judging from  the scenarios with a metal-poor stellar component,
there is no evidence for a significantly enhanced SFR  50-1000 Myr 
ago, as derived from the observed clusters. However, if we set
$Z_{\rm stars} \ge 0.008$, the cluster SFR remains high over the past  Gyr although
all the peaks at $t<100$ Myr remain. What happens is that many clusters
 are found to be younger if the metallicity is increased. Figure 
\ref{sfhlong} shows the cluster SFR over the  past 15 Gyr for 
three different metallicities ($Z_{\rm stars} =0.001$, 0.008 and 0.020). 
In addition, we
show the result from a P\'EGASE.2 evolutionary scenario. The present
starburst is still evident in the first bin ($0 < t \le 500$ Myr)
but has decreased in amplitude since the SFR has been averaged over
a longer time-scale. For the $Z_{\rm stars} =0.001$ scenario, a prominent
peak is found at 1.5 Gyr, and significant cluster formation
occurred also between 2 and 5 Gyr ago, which was found also by \"OBR98. 
Assuming that all GCs older than  10 Gyr formed during a short time 
scale ($\tau < 1$ Gyr) would produce a peak comparable to that found
at 1.5 Gyr. The event 1.5 Myrs ago must have been an impressive 
 starburst.

Bergvall \& \"Ostlin (\cite{bergvallostlin}) reported that the underlying host galaxy may
be significantly more metal-rich than the nebular gas. In this case
one might suspect that  older clusters could be
metal-rich. For higher metallicities, the 1.5 Gyr feature remain but decrease in 
amplitude. The 14
Gyr peak, on the other hand, becomes stronger. This is  due to the higher $M/L$ at high ages
in these scenarios, making  both the  mass detection limit and the 
 masses of individual
clusters higher.   
The unknown  metallicity of older clusters and the small number of
objects in each age bin presents the most important limitation
for estimating the past SFR from clusters.

\subsection{The cluster mass vs total galaxy mass}
\label{clustervstotal}

Summing up the total cluster SFR over 15 Gyrs we get $\langle SFR \rangle
= 0.02 M_\odot$/yr, or a total mass formed in clusters of  $\sim 3 \times 
10^8M_\odot$. 
The total stellar mass of this galaxy is estimated to be 
 $\sim 4 \times 10^9M_\odot$ (\"Ostlin et al. \cite{ostlin2001}, Bergvall \& 
\"Ostlin \cite{bergvallostlin}). Hence the clusters would be responsible for
almost 10\% of the total star formation. This is orders of magnitude higher
than for most galaxies, for which the typical value is 0.3\%
(McLaughlin \cite{mclaughlin99}, \cite{mclaughlin}). For instance, the total mass of the
Milky Way globular cluster system is estimated to be on the order of
 $\sim 6 \times 10^7M_\odot$. As noted in Sect. 4.1, if the actual IMF 
is more similar to Scalo98 than Salpeter, we will overestimate the
cluster masses. In addition, we count the total consumed gas mass 
(including the gas returned from stellar winds and explosions), but these
two effects combined should not be greater than a factor of 2. Old clusters 
may have suffered loss of low mass stars through dynamical evolution, but also
this effect should be not greater than a factor of 2 in view of the mass to
light ratios for Galactic globular clusters (Ashman \& Zepf \cite{ashmanzepf}).
Taking also the mass of gas $M(\ion{H}{i})=3\times 10^9 M_\odot$ into account we
are still an order of magnitude above  0.3\%.
If we count only the actually detected clusters 
 we still get a total cluster mass of  $9 \times 10^7 M_\odot$
for $Z_{\rm stars} = 0.001$ ($\ge 7 \times 10^7$ for any $Z_{\rm stars}$),
or about one third of the total implied cluster mass. This would still
give a mass fraction $\gg 0.3$\% if accounting for the possible systematic 
effects mentioned above.

The conclusion 
that this galaxy is, and has been over a cosmological time-scale, a very
efficient cluster former is unavoidable. The open question is why?
Tashiro and Nishi (\cite{tashiro}) suggest that the formation of gravitationally
bound clusters may be favoured in low metallicity environments.

Bergvall \& \"Ostlin (\cite{bergvallostlin}) found that the stellar metallicity 
of the host galaxy is higher than that of the ISM, 
which could be explained if a metal-poor gas-rich dwarf galaxy 
or an \ion{H}{i} cloud has been accreted. They speculate that the unusually
high metallicity, in view of the low luminosity, arise because the galaxy is 
rich in dark matter (see Sect \ref{halo}), which has enabled it to retain all 
its metals. If this is true,  the total mass fraction (including dark matter) 
of globular and akin clusters may not be 
higher than in other galaxies. Another possibility if one postulates a universal
cluster formation efficiency is that for some reason the survival chances
are greater in ESO\,338-IG04, but then this should be a general trend among
dwarf galaxies.

\section{Spatial propagation of the starburst}

\begin{figure*}
\resizebox{\hsize}{!}{\includegraphics{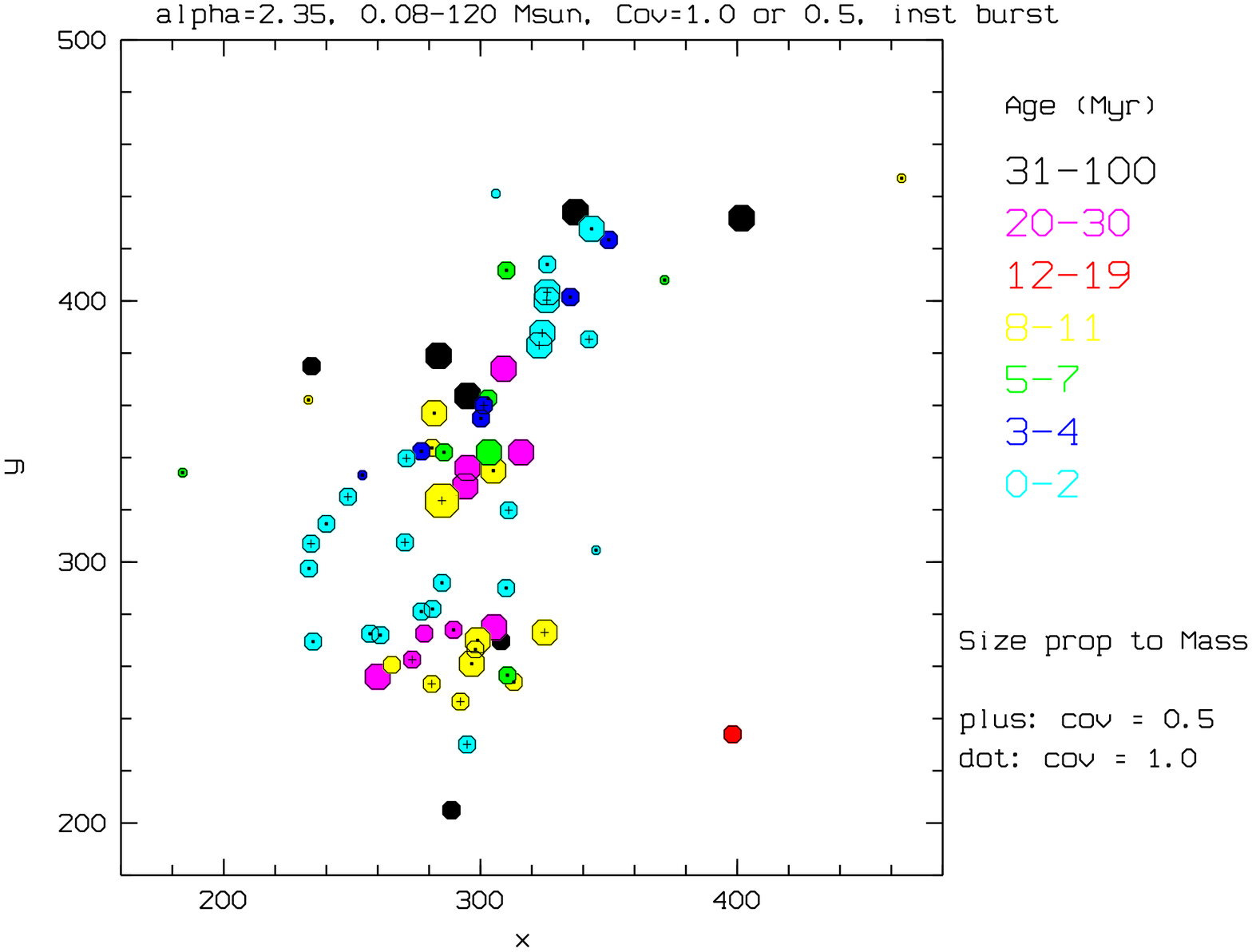}}
\caption{Spatial distribution of young clusters using best fitting $E(B-V)$
and Salpeter IMF with mass range 0.08 to 120 $M_\odot$, and metallicity
$Z_{\mathrm{stars}}=0.001$ and $Z_{\mathrm{gas}}=0.002$. The covering
factor was allowed to vary between 0.5 (+ signs) and 1.0 (dots). The size of 
each symbol is proportional to the logarithm of the mass of a cluster and 
the colour depends on the age. 
The x- and y- coordinates correspond to pixel numbers in the Planetary Camera  
and the region showed is $14.6\arcsec \times 14.6\arcsec$, or
$2.6 \times 2.6$ kpc.}
\label{pattern}
\end{figure*}

The distribution of young clusters can be used to investigate
if propagating star formation  have occurred. In Fig. 
\ref{pattern} we show the spatial age pattern for young clusters
for a standard metal-poor Salpeter scenario (our favoured
one for young objects) with a varying covering factor. 
There are indeed some striking patterns present in this image. 

 The youngest objects ($<3$ Myr), coded in cyan, tend to
prefer a mix of covering factors. They 
are preferentially found in a ring like structure centred on X=270, 
Y=305 (capital X and Y refer to Planetary Camera pixel values,
each pixel has a size of 8.3 pc, see Fig. \ref{pattern}) and in a group
near X=320, Y=400. The former structure have in general a best
fitting covering factor of 1.0, and the latter 0.5.

Slightly older objects ($3 \le$ Age $ <8$ Myr, coded in blue), are found in the
latter group and also near the starburst centre (X=290, Y=340). Objects
with ages 8 to 30 Myr  are found  in the
centre and in a group near (X=300, Y=260). Older objects (30-100) Myr
are not strongly clustered, but are preferentially found in the
upper half.

The ring-like structure of very young ($<3$ Myr) objects have a major axis diameter
of  0.8 kpc. A velocity of 1 km/s correspond
to 1 pc/Myr. The H$\alpha$ velocity dispersion is $\sigma_{{\rm H}\alpha} = 43$ km/s
(\"Ostlin et al. 2001) implying a turbulent half ring crossing time 
of 10 Myr. Hence the coeval formation of this ring  can 
not have been triggered by a wave travelling at
the ISM turbulent velocity, but
perhaps by a supernovae driven wind. The grouping of very young
objects in the upper part is more compact and consistent with a
coeval formation if random motions of 0.1 kpc are considered.
 To the right of this grouping several diffuse  emission-line
regions without compact clusters are present (see Fig. \ref{image}),
which may possibly be the young sites of star formation. 
The $\sim 10$ Myr (yellow) grouping in the lower part  (X=300, Y=260)
has a similar dimension. Their age is consistent with a localised 
formation followed by  diffusion by random motions. In
this region we find an interesting boundary between $\sim 10$
and $\sim 25$ Myr old objects, although one can not exclude that
a dust filament is responsible for the apparent ridge of  $\sim 25$ Myr
objects. 
The centre (X=290, Y=340) is close to the centre of the outer isophotes and
could therefore represent the central potential of the galaxy. Here
cluster formation appears to have been continuing over the last
40 Myr. The present starburst may have started 
in the region slightly west (up in Fig. \ref{pattern}) of the centre
where almost all of the sources older than 30 Myr are located. 
A possible scenario is that the upper and lower 8 to 30 Myr
groups represent approaching complexes that  have triggered
the young shell-like structure. Indeed, the H$\alpha$
velocity field show that the eastern  part of the galaxy 
(down-left in Fig. \ref{pattern})  may be falling in towards
the centre (\"Ostlin et al. 2001). Moreover, the H$\alpha$ velocity 
dispersion is higher east of this region ($\sigma_{{\rm H}\alpha} = 70$ km/s)
and there are suggestions of a perpendicular velocity component.
Further out towards east (outside the Planetary Camera FOV
 shown in Fig. \ref{image}) we find some diffuse 
emission-line regions and clusters\footnote{Some blueish clusters in this
tail ( \"Ostlin et al. 2001, Bergvall \& \"Ostlin 2002) are visible
in the Wide Field chip 3, but their photometry is too uncertain to
assess their ages.}.

Although projection effects and the
velocities of individual clusters are not known, our results
provide evidence for the spatial propagation of the starburst 
over the last 30 Myr. Supersonic triggering appears
to be necessary to explain some of the youngest features.

The star formation pattern and the conclusions remain unaffected if 
a fixed $E(B-V)=0.0$ or 0.05, or a fixed covering factor, is adopted.
The same applies to changing the IMF: ages  would change slightly, 
but not  the pattern.
If massive star formation would be associated with a time-lag, as 
discussed above, most features would remain, e.g. the young ring,
but it would be a few Myrs older, but still too young for a sound
wave to have triggered their formation. Some of the $\sim$10 Myr
objects (in yellow) would now rather be just a few Myr.

\subsection{Spatial distribution of old globular clusters}
\label{halo}

\begin{figure}
\resizebox{\hsize}{!}{\includegraphics{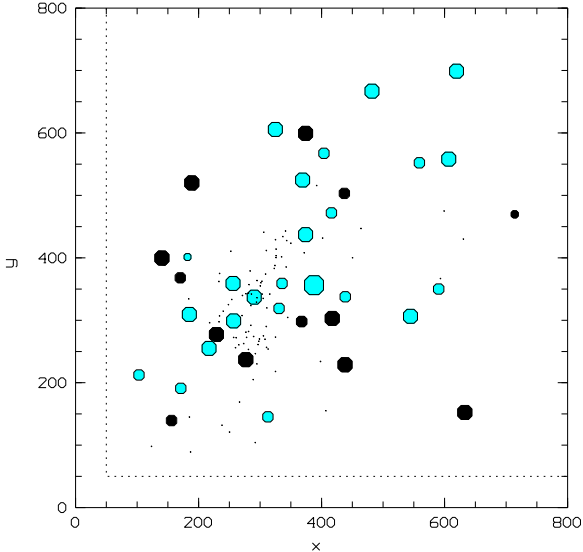}}
\caption{Spatial distribution of old (age $> 5$ Gyr, filled black 
symbols) and intermediate age ($0.8 <$ age $\le 5$ Gyr, open shaded symbols)
clusters with size proportional to mass. The small dots younger ($\le 0.8$ Gyr) 
objects. The axes show the spatial 
coordinates, in units of PC pixels.  The region delimited by a dashed line ($X < 50, Y<50$)
lies at the border to the wide field camera CCDs  and is subject to
vignetting, and therefore not included in our analysis.}
\label{oldpattern}
\end{figure}

For  old ($\ge 0.8$ Gyr) clusters the time-scale is too long for the
position of individual clusters to have any correspondence with their 
place of birth. Instead, we examine the spatial distribution of the 
ensemble of old clusters (Fig. \ref{oldpattern}). 
Intermediate age ($0.8 \le$ age $<5$ Gyr) clusters (open shaded symbols in 
Fig. \ref{oldpattern}) have a flattened distribution with minor to major 
axis ratio $b/a \approx 0.5$, with the same position angle as the galaxy 
as a whole. The isophotes of ESO\,338-04 
have an axis ratio $b/a = 0.48$ (Bergvall \& \"Ostlin \cite{bergvallostlin}).
The colours of the host galaxy at radii sufficiently large that the starburst
contribution is negligible, indicate an age in the range 2-8 Gyr and a 
relatively high stellar metallicity. 
That the intermediate age globular cluster system has an identical flattening
suggests a common origin or, if the lower age of the clusters is taken at face 
value, that these formed in the existing gravitational potential of the
 host galaxy. If a higher metallicity is adopted for the clusters, the age 
difference would  increase.

On the contrary, clusters older than 5 Gyr  (filled black
symbols in Fig. \ref{oldpattern}) show no apparent flattening 
($b/a \approx 1.0$) suggesting the presence of an old spherical 
halo, and that they are indeed metal-poor.
 The number of objects is  too small to study the
radial distribution in detail, but the effective radius (containing half the 
number of  $>$5 Gyr clusters) is $\sim 1.4$ kpc. If a Gaussian LF
similar to the one for Galactic GCs ($M_v^{peak}=-7.2, \sigma=1.3$ mag) is assumed, 
the total number (including non detected ones) of  old halo clusters 
should be $\sim$50.  
For a typical GC richness of spherical systems ($T = 5$, see Ashman \& Zepf 
\cite{ashmanzepf} for a definition of $T$) 50 GCs would then imply a mass of the 
halo of $10^{10} M_\odot$. This thought experiment supports the idea of a dark 
matter-rich halo discussed in Sect. \ref{clustervstotal}

\section{Discussion -- Star formation and dusty molecular clouds}

Star formation in our Galaxy and elsewhere is associated with dusty 
giant molecular clouds. Massive starbursts like M82 and Arp220 
are reddened by several tens of magnitudes of optical extinction
(Genzel et al. \cite{genzel}), and a large fraction of the UV power 
is absorbed and re-radiated in the infrared. However, the
extinction inferred from young star clusters in M82 is
only of the order of $A_V=1$ (de Grijs et al. \cite{degrijs}) 
compared to the  50 magnitudes derived from IR data (F\"orster 
Schreiber et al. \cite{forster}). In the Antennae, the youngest 
clusters have high extinction $A_V = 3$ to 5 (Zhang 
et al \cite{Zhang et al.}), and are spatially correlated with the IR 
emission, whereas the nuclei have $A_V\sim50$. Obviously, optical selection 
systematically picks out sources with lower than average extinction.

The situation in ESO\,338-04 is quite different. 
Looking at  Fig. \ref{xubv} and \ref{galvspeg} 
we see no tendency for the youngest objects to be  reddened.
A few sources lie to the upper right compared to the locus of 
the evolutionary sequences at the lowest age, as if they were reddened by 
$E(B-V) \le 0.25$. However, there are more sources to the down-left, 
and the observations are  consistent
with a dust free population plus random effects (integer number
of massive stars, and small variations is gas filling and covering 
factors) plus photometric errors. Although our sample
 is biased towards  optically bright
 clusters it is nevertheless surprising that 
we do not see any young sources just about to come out of their dusty 
cocoons in which they, according to common wisdom, should have formed. 
Note that there are nebular emission dominated regions without associated
point sources west of the centre, which may represent young 
objects, still partly embedded.

The  lifetime of Galactic giant molecular clouds against 
photo- dissociation and ionisation is on the order of 30 Myr once massive 
star formation commence (Williams and McKee \cite{williams}). If we
insist that star formation is associated with dusty molecular clouds,
then the lack of young reddened sources suggest that 
in metal-poor starbursts with high ISM turbulence (\"Ostlin et al. 
\cite{ostlin2001}) and high star formation efficiency, this time scale 
is much shorter. ESO\,338-04 is a hot IRAS source ($S_{60}/S_{100}=1.7$)
but the FIR luminosity is modest $L_{\rm IR}=5.8\times10^9 L_\odot$
and the IR to blue luminosity, $L_{\rm IR}/L_B = 1.0$ (solar units),
is very low with respect to its  $f_{60}/f_{100}$ ratio (Bergvall et al. 
\cite{bergvalletal}, Dulzin-Hacyan et al.  1988). The derived dust mass
is only of the order of $10^5 M_\odot$ (see also Calzetti et al \cite{calzetti95}). 
This implies small amounts of dust heated to high temperatures, 
and fits well with the short time-scale scenario.

A short time-scale for molecular cloud destruction would  be consistent 
with the difficulty to find CO emission in metal-poor  dwarf starbursts
(Kunth \& 
\"Ostlin  \cite{kunthostlin}) since not only the CO 
to H$_2$ conversion factor may be low, 
but the gas would stay in molecular form a shorter time. 
For ESO\,338-04, $L_{CO(1-0)} < 7\times10^{27}$W, so $\log(L_{CO}/L_{IR}) 
< -8.4$ (Bergvall, unpublished), implying $M_{{\rm H}_2} \le 7 \times 10^6 M_\odot $ if a 
standard CO to ${\rm H}_2$ conversion factor is used. As the mass of young
($\le 10$ Myr) stars (in clusters and diffuse) is  $\sim 10^7 
M_\odot$, and star formation efficiency must be well below 100\%,
the molecular gas dissociation time scale should be
a few Myr at most.

While we see several very young ($\sim$1 Myr) clusters, 
Fig. \ref{sfhshort} shows  that the global star formation 
intensity may have passed its peak. 
This combined with the tendency to pick out sources with low extinction
suggest that some star formation could be hidden after all. Of course
 we may happen to observe this galaxy just as the last reservoirs of dusty 
molecular gas are depleted, but we would
still need to postulate a short dust destruction time scale.

If the time scale for molecular cloud destruction is unusually short,
it is possible that a young cluster would still not have reached 
its peak UV luminosity when becoming naked. This could give us an additional 
selection effect to  account for the lack of young reddened clusters.

The cluster richness implies a high cluster and star formation efficiency 
(see Sect. 8.3), which could be a natural cause for the short dust 
destruction time-scales since the UV radiation density in the ISM is high. 
It is surprising that the star formation efficiency can be so high in a galaxy
with a metal-poor ISM and little dust.

\section{Conclusions}

Using multicolour photometry from 2000 to 8000\AA, we have analysed the temporal
and spatial evolution of  star clusters and the starburst in the Luminous
Blue Compact Galaxy ESO\,338-04. Our main results are:

-- Using an extensive set of spectral evolutionary synthesis scenarios
   and Monte-Sarlo simulations we
   conclude that the overall age distribution found is very robust. Massive star
   clusters with ages from 1 Myr up to the maximum age considered in the models (15 Gyr) are present in this galaxy.
 
-- A careful treatment of the nebular emission is crucial for modelling the
   properties of young star clusters.

-- In accordance with previous studies, we find the starburst in
   ESO\,338-04 to be almost unaffected by dust extinction. There
   is no evidence for a large population of young  sources about to come
   out of their dusty cocoons. We speculate that this can be explained by a short
   time-scale, $\le 1$ Myr, for destruction of dusty molecular clouds. 

-- The cluster formation history have been used to constrain the overall star 
   formation history in this galaxy. The present starburst began 40-50 Myr ago
   and is still active. There are indications that the SFR has varied
   during time, and that it has already passed its peak.

-- Young clusters are responsible for 30 to 40\% of the star formation and  the
   starburst luminosity output in the UV, which is a high number, even for starbursts.

--   Seen over a cosmological
   time, this galaxy has been a remarkably efficient cluster former. The reason
   for this remains an enigma. The 
   intermediate age population has a flattened distribution with the same 
   flattening as the host galaxy. The old clusters have a spherical distribution.

-- The observations are well reproduced by a Salpeter IMF ($\alpha=2.35$) 
   with an upper mass limit $M_{\rm up} \ge 60 M_\odot$. A steeper IMF ($\alpha=2.85$) 
   cannot reproduce the data. A flatter IMF ($\alpha=1.35$)  cannot be excluded but
   requires additional constraints to provide a good fit. An ISM metallicity close to the 
   spectroscopic value is strongly favoured. Our results  suggest that 
   the formation time-scale for an individual cluster is  at maximum a few Myrs.

-- We find evidence for spatially correlated formation of
   groups of clusters, some requiring supersonic triggering velocities, e.g.
   from supernovae  winds.

\begin{acknowledgements}
This work was supported by the  Swedish  Research
Council and  the Swedish National Space Board. 

\end{acknowledgements}


\begin{thebibliography}{}

\bibitem[1998]{ashmanzepf} Ashman K.M.,  Zepf S.E., 1998, Globular Cluster Systems, 
Cambridge University Press 

\bibitem[1985]{bergvall}  Bergvall N., 1985, A\&A 146, 269

\bibitem[2002]{bergvallostlin} Bergvall N., \"Ostlin G., 2002, A\&A 390, 891 
     
\bibitem[2000]{bergvalletal} Bergvall N., Masegosa J., \"Ostlin G., Cernicharo J., 2000, A\&A 359, 41 

\bibitem[2002]{biretta} Biretta, Lubin., et al. 2002, WFPC2 Instrument Handbook, Version 7.0 (Baltimore: STScI).

\bibitem[1991]{brodie} Brodie, J.P., Huchra, J.P., 1991, ApJ 379, 157

\bibitem[2002]{buat} Buat V., Burgarella D., Deharveng J.M., Kunth D.,  2002, A\&A 393, 33

\bibitem[1994]{calzetti} Calzetti D., Kinney A.L., Storchi-Bergmann T., 1994, ApJ 429, 582

\bibitem[1995]{calzetti95} Calzetti D., Bohlin R.C., Kinney A.L., Storchi-Bergmann T., Heckman T.,
1995, ApJ 443, 136

\bibitem[1996]{charlotetal}
 Charlot S., Worthey G., Bressan A., 1996, ApJ 457, 625

\bibitem[1987]{Clegg & Middlemass}
 Clegg, R.E.S, \& Middlemass D. 1987, MNRAS 228, 759

\bibitem[2001]{degrijs}
De Grijs R., O'Connel R.W., Gallagher J.S., 2001, AJ 121, 768 

\bibitem[dulzin]{1988} Dulzin-Hacyan D., Moles M., Masegosa J., 1988, A\&A 206, 95

\bibitem[2001]{fall} Fall S.M, Zhang Q., 2001 ApJ 561, 751

\bibitem[1996]{Ferland}
 Ferland, G.J. 1996, Hazy, a Brief Introduction to Cloudy. University of Kentucky Department of Physics and Astronomy Internal Report.

\bibitem[2001]{forster} F\"orster Schreiber N.M, Genzel R., 
Lutz D., Kunze D., Sternberg A., 2001, ApJ 552, 544

\bibitem[1999]{pegase}
 Fioc M., \& Rocca-Volmerange B., 1999, astro-ph/9912179

\bibitem[1998]{genzel} Genzel R., Lutz D., et al.
1998, ApJ 498, 579


\bibitem[1997]{guzman} Guzm\'an R., Gallego J., Koo D.C., et al.
1997, ApJ 489, 559  

\bibitem[2002]{hopkins} Hopkins A.M., Schulte-Ladbeck R.E., Drozdovsky I.O., 2002, AJ 124, 862

\bibitem[1998]{iye} Iye M., Ulrich M.-H., Peimbert M., 1987, A\&A 186, 84

\bibitem[1989]{kassim} Kassim N.E., Weiler K.W., Erickson W.C., Wilson T.L., 1989, ApJ 338, 152

\bibitem[1998]{kennicutt} Kennicutt R.C., 1998, in ``The Stellar Initial Mass Function'' eds. Gilmore G, Howell D., ASP Conf. Ser. 142, 1

\bibitem[2002]{kroupa} Kroupa P., 2002, Science 295, 82

\bibitem[2000]{kunthostlin} Kunth D., \"Ostlin G., 2000, A\&ARev 10,1

\bibitem[2000]{lancon} Lan\c{c}on A., Mouhcine M., 2000, in  'Massive Stellar Clusters', 
eds. A. Lan\c{c}on and C. Boily, ASP Conf. Series 211,  63

\bibitem[2000]{lefevre} Le F\`evre O., Abraham R., Lilly S.J., et al.,
2000, MNRAS, 311, 565

\bibitem[1998]{leitherer} Leitherer C., 1998, in ``The Stellar Initial Mass Function'' eds. Gilmore G, Howell D., ASP Conf. Ser. 142, 61

\bibitem[1998]{Lejeune et al.}
 Lejeune, T., Cuisinier, F., \& Buser, R. 1998, A\&AS 130, 65L

\bibitem[1996]{lilly} Lilly S.J., Le F\`evre O., Hammer F., Crampton D., 1996, Apj 460, L1 

\bibitem[1996]{madau} Madau P., Ferguson H.C., Dickinson M.E., Giavalisco M., Steidel C.C., Fruchter A., 1996, MNRAS 283, 1388

\bibitem[2001]{maiz} Ma\'iz-Appel\'aniz J., 2001, ApJ 563, 151

\bibitem[1999]{mas-hesse} Mas-Hesse J.M., Kunth D., 1999, A\&A 349, 765

\bibitem[1998]{mh} Massey P., Hunter D.A., 1998, ApJ 493, 180

\bibitem[2002]{mckee} McKee C.F., Tan J.C., 2002, Nature 416, 59 

\bibitem[2000]{mclaughlin} McLaughlin D., 2000, in  'Massive Stellar Clusters', eds. A. Lan\c{c}on 
and C. Boily, ASP Conf. Series 211,  63

\bibitem[1999]{mclaughlin99} McLaughlin D., 1999, AJ 117, 2398

\bibitem[1995]{meurer} Meurer  G.R., Heckman T.M., Leitherer C., Kinney A., 
Robert C., Garnett D.R., 1995, AJ 110, 2665 

\bibitem[1999]{meurer99} Meurer  G.R., Heckman T.M., Calzetti D., ApJ 521, 64

\bibitem[1997]{meylanheggie} Meylan G., Heggie D.C., 1997, A\&ARev 8, 1

\bibitem[1982]{millerscalo} Miller G., Scalo J., 1982 ApJ 263, 259

\bibitem[1998]{ostlin1998} \"Ostlin G., Bergvall N., R\"onnback J.,
        1998, A\&A, 335, 85 (\"OBR98)

\bibitem[2000]{ostlin2000} \"Ostlin G., 2000, in  'Massive Stellar Clusters', eds. A. Lan\c{c}on 
and C. Boily, ASP Conf. Series 211,  63

\bibitem[2001]{ostlin2001} \"Ostlin G., Amram P., Bergvall N., Masegosa J., 
Boulesteix J., Marquez I., 2001,  A\&A 374, 800

\bibitem[2000]{pandey} Pandey A.K., Ogura K., Sekiguchi K., 2000, PASJ 52, 847 

\bibitem[1998]{pascarelle} Pascarelle S.M., Windhorst R.A., Keel W.C., 1998, AJ 116, 2659 

\bibitem[2000]{raimann}Raimann D., Storchi-Bergmann T., Bica E.,
 Melnick J., Schmitt H., 2000, MNRAS 316, 559

\bibitem [1955]{salpeter} Salpeter E., 1955, ApJ 121, 161

\bibitem[1998]{Scalo}
Scalo, J.M. 1998, in ``The Stellar Initial Mass Function'', ed. G. Gilmore, \& D. Howell, ASP Conf. Ser. 142, p. 201 (Scalo98)

\bibitem[2001]{schulte} Schulte-Ladbeck R.E., Hopp U., Greggio L., Crone M.M., Drozdovsky I.O., 
2001, AJ 121, 3007

\bibitem[1972]{ss} Searle L., Sargent W.L.W., 1972, ApJ 173, 25  

\bibitem[1979]{seaton} Seaton M.J., 1979, MNRAS 187, 73

\bibitem[1998]{serabyn} Serabyn E., Shupe D., Figer D.F., 1998, Nature 394, 448

\bibitem[2000]{tashiro} Tashiro M., Nishi R., 2000, ApJ 536, 277 

\bibitem[2002]{vacca} Vacca W.D., Johnson K.E., Conti P.S. 2002, AJ 123, 772

\bibitem[2002]{whitmore_rev} Whitmore B.C., in ``A Decade of Hubble Space Telescope Science''
Eds. Livio M., Noll K., Stiavelli M., Cambridge University Press, in press. (astro-ph/0012546) 

\bibitem[2002]{whitmore&zhang} Whitmore B.C., Zhang Q., 2002, AJ 124, 1418


\bibitem[1999]{whitmoreetal} Whitmore B.C., Zhang Q., Leitherer C., Fall S.M., Schweizer F., Miller B.W., 1999, AJ 118, 1551


\bibitem[1996]{hdf} Williams R.E., Blacker B., Dickinson M., et al., 1996, AJ 112, 1335 

\bibitem[1997]{williams}Williams J.P, McKee C.F., 1997, Apj 476, 166 


\bibitem[2001]{Zackrisson et al.}
Zackrisson, E., Bergvall, N., Olofsson, K., \& Siebert, A. 2001, A\&A 375, 814

\bibitem[2001]{Zepf et al.} Zepf  S.E., Ashman K.M., English J., Freeman K.C., Sharples R.M.
1999, AJ 118, 752

\bibitem[2001]{Zhang et al.}
Zhang Q., Fall S.M., Whitmore B.C., 2001, ApJ 561, 727

\end{thebibliography}
\end{document}